\pdfoutput=1
\documentclass[a4paper,11pt]{article}

\usepackage{jcappub}
\usepackage{graphicx}
\usepackage{amsmath,amsthm,latexsym,amssymb,amsfonts}
\usepackage{hyperref}
\usepackage[usenames,dvipsnames]{xcolor}

\newcommand{\td}{{\rm d}}
\newcommand{\vect}[1]{\boldsymbol{#1}}
\DeclareMathOperator{\artanh}{artanh}
\DeclareMathOperator{\arcoth}{arcoth}
\newcommand{\be}{\begin{equation}}
\newcommand{\ee}{\end{equation}}
\newcommand{\bea}{\begin{equation}\begin{aligned}}
\newcommand{\eea}{\end{aligned}\end{equation}}
\def\lsim{\mathrel{\raise.3ex\hbox{$<$\kern-.75em\lower1ex\hbox{$\sim$}}}}
\def\gsim{\mathrel{\raise.3ex\hbox{$>$\kern-.75em\lower1ex\hbox{$\sim$}}}}

\title{Microlensing of gravitational waves by dark matter structures}

\author[1]{Malcolm~Fairbairn,}
\author[2,3]{Juan~Urrutia}
\author[2,4,5]{and Ville~Vaskonen}
\affiliation[1]{Department of Physics, King's College London, Strand, WC2R 2LS London, UK}
\affiliation[2]{National Institute of Chemical Physics and Biophysics, R\"avala 10, Tallinn, Estonia}
\affiliation[3]{Division of Physics, Tallinn University of Technology, Akadeemia tee 21, 12618 Tallinn, Estonia}
\affiliation[4]{Dipartimento di Fisica e Astronomia, Universit\`a degli Studi di Padova, Via Marzolo 8, 35131 Padova, Italy}
\affiliation[5]{Istituto Nazionale di Fisica Nucleare, Sezione di Padova, Via Marzolo 8, 35131 Padova, Italy}
\emailAdd{malcolm.fairbairn@kcl.ac.uk}
\emailAdd{juan.urrutia@kbfi.ee}
\emailAdd{ville.vaskonen@kbfi.ee}

\abstract{Gravitational lensing of gravitational waves provides a potential new probe of dark matter structures. In this work, we consider the microlensing effect on gravitational wave signals from black hole binaries induced by low-mass dark matter halos that do not retain enough baryonic matter to hold stars. We clarify systematically when this microlensing effect is relevant and study in detail its detectability by future gravitational wave observatories. We consider lensing by cold dark matter halos and by solitonic cores that reside in fuzzy dark matter halos. Our results show that although the effect can be detectable at relatively large impact parameters, the probability of detecting such lensed events is low. In particular, we find that the expected number of events lensed by cold dark matter halos is $\mathcal{O}(0.01)$ per year for BBO and the expected number of events lensed by solitonic cores inside fuzzy dark matter halos is $\mathcal{O}(0.01)$ per year for ET. In the case that a significant fraction of dark matter consists of $\mathcal{O}(100 M_\odot)$ objects that are relatively compact, $R < \mathcal{O}(0.1\,{\rm pc})$, we show that the expected number of lensed events per year for ET can be very large, $\mathcal{O}(1000)$.}

\begin{document}

\maketitle

\section{Introduction}

Gravitational wave (GW) observations have opened a new window into the Universe and may uncover deep questions in modern physics. With the LIGO-Virgo observations we have, for example, learned about the black hole (BH) population~\cite{LIGOScientific:2021psn}, tested general relativity in the strong regime~\cite{LIGOScientific:2021sio}, constrained the graviton mass~\cite{LIGOScientific:2021sio}, the speed of GWs~\cite{LIGOScientific:2021sio} and the nuclear matter equation of state~\cite{LIGOScientific:2018cki}. Cosmology with GWs is also promising. Already LIGO-Virgo has been able to measure the Hubble constant~\cite{LIGOScientific:2021aug} and put constraints on GW backgrounds~\cite{LIGOScientific:2021nrg,KAGRA:2021kbb,Romero-Rodriguez:2021aws,Romero:2021kby}. The new more sensitive observatories, including ET~\cite{Punturo:2010zz}, CE~\cite{Reitze:2019iox}, AEDGE~\citep{AEDGE:2019nxb,Badurina:2021rgt}, LISA~\citep{LISA:2017pwj}, and BBO~\cite{Crowder:2005nr}, will in the near future expand the frequency band providing more precise measurements and potentially answer further questions.

One of the most puzzling aspects of cosmology is the elusive nature of dark matter (DM). In this regard, one of the ways that GW observatories may be capable of detecting DM is through the gravitational interaction of GWs with DM substructures. For example, the cold dark matter (CDM) paradigm predicts an abundance of light halos which may act as lenses for a GW passing nearby \cite{Diemand:2008in}.  Halos lighter than about $10^8 M_\odot$ do not retain enough baryonic matter to hold stars~\cite{Sawala:2014hqa} so their detection with conventional techniques is very challenging. Deviations from CDM, which appears to work well down to this mass scale, may appear at these lower masses.  

At these low mass scales, structure can be wiped out due to a large free streaming length in the early Universe caused by the DM having a non-zero kinetic energy (i.e. being warm) during structure formation~\cite{Bode:2000gq}.  Similarly, in ultralight DM scenarios, the Compton wavelength of DM can become comparable to astrophysical distances and the formation of halos with characteristic size below this length scale is suppressed in a way closely analogous to the suppression of structure on small scales in warm DM scenarios~\cite{Hui:2016ltb,Rogers:2020ltq}.

There are two other ways that dense dark objects created in fuzzy (ultra-light) DM scenarios are different from those created in traditional CDM scenarios which will be particularly relevant for this work - The first is the fact that above that suppression length/mass scale, the necessarily high occupation number of ultra-light DM in dense regions leads to the formation of solitonic cores at the center of dark halos \cite{Hui:2016ltb,Schive:2014dra,Marsh:2015wka}.  While such halos host almost constant density cores with lower central densities than might occur in CDM scenarios, the mass as a function of radius can be slightly higher close to the edge of the central solitonic region which can affect gravitational lensing.

The second is related to the particle physics of the situation in the early Universe - In models where the ultra light DM is a Nambu-Goldstone boson of a broken U(1) symmetry, as is the case for axion-like particles if the symmetry is broken after inflation, the Kibble Mechanism results in large differences in field expectation values from Hubble volume to Hubble volume.  This scaling with the Hubble volume continues until the resulting energy differences are converted into dense halos at the moment where the mass of the boson becomes comparable to the Hubble expansion rate, and these halos posses larger density contrasts which start to grow earlier than conventional CDM structures emerging from small adiabatic fluctuations on larger scales \cite{Kolb:1993zz,Kolb:1994fi,Ellis:2022grh}.  

The evolution of these halos and their subsequent mergers is still a subject of debate, current thought being that they form a spectrum around the characteristic mass imprinted at the stage where the scaling halts \cite{Fairbairn:2017sil,Ellis:2022grh}.   Across a broad mass range, the existence of such cores can be probed through optical microlensing~\cite{Fairbairn:2017dmf,Fairbairn:2017sil}.  

What is clear is that studying these smaller objects may shed light on the nature of DM and the community is advocating for the importance of finding new ways to probe DM structures at these lower masses by observing the disruption of stellar streams~\cite{Banik:2018pal,Banik:2019cza} and the detection of smaller halos through lensing of photons~\cite{Gilman:2019vca,Fairbairn:2022gar,Nightingale:2022bhh}.  In this work, we aim to investigate the possibility of probing such halos using their lensing of GWs.

The theory of gravitational lensing of GWs was originally addressed in~\cite{Lawrence:1971hx,Ohanian:1974ys}, and further developed in~\cite{Wang:1996as,Nakamura:1997sw,Zakharov:2002bq,Takahashi:2003ix,Seto:2003iw,Matsunaga:2006uc,Sereno:2010dr}. The interest in this topic has grown in the recent years~\cite{Piorkowska:2013eww, Biesiada:2014kwa, Ding:2015uha, Dai:2016igl, Smith:2017mqu, Ng:2017yiu, Jung:2017flg, Liao:2017ioi, Li:2018prc, Oguri:2018muv, Congedo:2018wfn, Dai:2018enj, Broadhurst:2018saj, Broadhurst:2019ijv, Cao:2019kgn, Cao:2020sky, Li:2019rns, Oguri:2020ldf, Hannuksela:2020xor, Liao:2020hnx, Piorkowska-Kurpas:2020rfy, Urrutia:2021qak, Diego:2021fyd, Dalang:2021qhu, Choi:2021bkx, Gao:2021sxw, Guo:2022dre, Cremonese:2021ahz, Mpetha:2022xqo, Caliskan:2022hbu, Zhou:2022yeo} although the searches performed by the LIGO-Virgo collaboration have not shown evidence for lensing signatures~\cite{LIGOScientific:2021izm}. In~\cite{Urrutia:2021qak} the non-observation of microlensing effects by LIGO-Virgo was used to derive a new constraint on the abundance of compact DM objects. The major motivation for the lensing studies, however, arises from the proposed new GW observatories. These observatories will detect orders of magnitude more events than LIGO-Virgo, in a wider range of masses and up to much longer distances, and, consequently, several lensed events are expected.

In this paper, we focus on the prospects of detecting microlensing imprints caused by light, $M_{\rm v} < 10^8 M_\odot$, DM halos in the GW signals from black hole binaries. Similar studies done recently in~\cite{Choi:2021bkx,Guo:2022dre} have shown optimistic results with several detectable lensed events per year even in the traditional CDM scenarios. Our work introduces several improvements over the previous studies. First, we clarify in a systematic way what are the conditions under which microlensing effects are relevant. Second, we considered realistic profiles for the lenses, including for the first time the effect of a constant density profile, which is relevant for the fuzzy DM cores. Moreover, instead of the often used singular isothermal sphere (SIS) profile, we consider the more realistic NFW and Einasto profiles for the CDM halos. Third, we perform a systematical matched filtering analysis of the detectability of the microlensing effect and calculate the probability at which a given source is lensed in a way that is detectable accounting for realistic halo mass functions. With these improvements, we find that the probability of seeing events that are microlensed by halos or their solitonic cores is in general very small for all future GW observatories. This is a new and important result for the planning of new gravitational wave observatories.

\section{Lensing by dark matter structures}

%\begin{figure}
%    \centering    \includegraphics[width=0.76\textwidth]{lenssystem.pdf}
%    \caption{A schematic picture of the source-lens-detector system. The source is off by $\vect{\eta}$ from the line of sight to the lens and $\vect{\xi}$ defines where the path crosses the lens plane. The vectors $\vect{\eta}$ and $\vect{\xi}$ are not in general aligned.}
%    \label{fig:lensing}
%\end{figure}

%The setup we consider is illustrated in Fig.~\ref{fig:lensing}: 
Consider a binary emits GWs at the angular diameter distance $D_s$, that on their way to the detector pass by a DM structure at the angular diameter distance $D_l$. The DM structure curves the spacetime causing gravitational lensing of the signal. In this work, we consider microlensing effects that modify the waveform due to interference of the different paths that the GW takes around the DM structure.\footnote{For simplicity, we consider isolated spherical lenses. We note, however, that non-sphericities of the lens and larger structures surrounding the lens may affect the observed signal~\cite{1994A&A...284..285K,Diego:2019lcd,Diego:2019rzc}.} In the frequency domain the lensed GW signal is $\tilde\phi_L(f) = F(f) \tilde\phi(f)$, where $\tilde\phi(f)$ denotes the signal that would be seen without the lens and $F(f)$ is the amplification factor. The latter can be written in the thin-lens approximation as~\cite{schneider2012gravitational}
\be \label{eq:F}
    F(w,\vect{y}) = \frac{w}{2i \pi} \int \td^2 \vect{x} \,e^{i w T(\vect{x},\vect{y})} \,,
\ee
where the integral is over the lens plane. Using a characteristic length scale $\xi_0$ of the system, the dimensionless vectors $\vect{x}$ and $\vect{y}$ are defined as $\vect{x} \equiv \vect{\xi}/\xi_0$ and $\vect{y} \equiv D_l \vect{\eta}/(D_s \xi_0)$ and the dimensionless frequency $w$ and the dimensionless time delay function $T$ as
\be
    w \equiv \frac{(1+z_l) D_s}{D_l D_{ls}} \xi_0^2 2\pi f \,,\qquad 
    T(\vect{x},\vect{y}) \equiv \frac12 |\vect{x}-\vect{y}|^2 - \psi(\vect{x}) - \phi(\vect{y}) \,,
\ee
where $f$ is the signal frequency, and $D_{ls} = D_s - D_l (1+z_l)/(1+z_s)$ the angular diameter distance between the lens and the source.\footnote{The expression for $D_{ls}$ is justified since the Planck and BAO observations indicate a very low cosmic curvature~\cite{Planck:2018vyg}.} The vector $\vect{\eta}$ determines the displacement of the source from the line of sight to the lens, $\vect{y}$ is its dimensionless projection to the lens plane, and the vector $\vect{\xi}$ is in the lens plane and determines the point through which the path crosses it. We call $y \equiv |\vect{y}|$ the impact parameter. The function $\phi(\vect{y})$ is defined such that the minimum of $T(\vect{x},\vect{y})$ with respect to $\vect{x}$ is zero, $\min_{\vect{x}} T(\vect{x},\vect{y}) = 0$.\footnote{$\phi(\vect{y})$ gives only an overall phase factor for $F$ but ensures that the phase of the signal is same for all potentials.} 

The dimensionless deflection potential $\psi(\vect{x})$ is obtained from the gravitational potential $U(\vect{r})$ of the lens by integrating over the direction perpendicular to the lens plane. For a spherically symmetric lens potential, $\psi(x)$ can be integrated from~\cite{Keeton:2001ss}
\be \label{eq:psix}
\partial_x \psi(x) = \frac{4D_l D_{ls}}{\xi_0 ^2 D_s} \frac{M_{\rm cyl}(x)}{x}\, ,
\ee
where $x=|\vect{x}|$. The projected mass in the lens plane within radius $x$ is given by
\be
M_{\rm cyl}(x) = 2\pi \int _0^{x}dx'\int_{-\infty}^\infty \td z \,  x' \rho(x',z) \,,
\ee 
where $z$ denotes the direction perpendicular to the lens plane and $\rho(x,z)$ the mass density of the lens. The constant of integration from~\eqref{eq:psix} can be included in $\phi(y)$.

In the geometric optics limit, i.e. the limit which roughly holds when the wavelength of the signal is much smaller than the Schwarzschild radius of the lens, the amplification factor can be approximated by accounting only for the paths that extremize the time delay function, $\nabla_x T(\vect{x}_j,\vect{y}) = 0$ or equivalently $\vect{y} = \vect{x}_j - \nabla_x \psi(\vect{x_j})$. The amplification factor in this limit reduces to
\be \label{eq:amp_geo}
    F(w,\vect{y}) = \sum_{i} |\mu(\vect{x}_j)|^{1/2} \,e^{iw T(\vect{x}_j,\vect{y}) - i\pi a_j} \,.
\ee
The magnification of the extremal path that crosses the lens plane at $\vect{x} = \vect{x}_i$ is given by
\be \label{eq:amplification}
    \mu(\vect{x}_j) = \frac{1}{ \det\left[ \partial_a \partial_b T(\vect{x}_j,y) \right]} = \frac{1}{\det\left[  \delta_{ab} - \partial_a \partial_b \psi(\vect{x}_j) \right]} \,,
\ee
and $a_j$ is $0$, $1/2$ or $1$ depending on whether $\vect{x}_j$ is a minimum, saddle point or a maximum of $T$, respectively.

If the time difference between different paths is small, they will interfere at the detector. This happens if
\be \label{eq:interference}
\frac{t(\vect{x},\vect{y})}{f} \left|\frac{\td f}{\td t} \right| \ll 1 \,,
\ee
where $t(\vect{x},\vect{y}) = \omega T(\vect{x},\vect{y})/(2\pi f)$. As the largest contribution to the integral comes from the paths near the paths $\vect{x}_j$ that extremize the time delay function, we can estimate if the interference condition~\eqref{eq:interference} is satisfied by considering the differences between the paths $\vect{x}_j$. If the interference condition~\eqref{eq:interference} is not satisfied, we enter the strong lensing regime where two or more signals from the same source can be distinguished. 

%In this work we will focus on the microlensing case where Eq.~\eqref{eq:interference} is fulfilled since the analysis of the detectability of the lens in the strong lensing case is different. In the strong lensing case we should study whether we can confidently say that the two or more signals are lensed counterparts of the same event~\cite{Haris:2018vmn} whereas in the microlensing case we look for deformations of the unlensed signal (see Sec.~\ref{sec:analysis}). In the following we will show that the microlensing case is relevant for a broad range of lens and source masses. 

In this work we will focus on the microlensing case where Eq.~\eqref{eq:interference} is fulfilled. To study the detectability of the lens we look for deformations of the unlensed signal (see Sec.~\ref{sec:analysis}). Our analysis is therefore considerably different than in the strong lensing case where we should study whether we can confidently say that the two or more signals are lensed counterparts of the same event~\cite{Haris:2018vmn}. The prospects of seeing strongly lensed GW signals have been studied e.g. in Refs.~\cite{Piorkowska:2013eww,Biesiada:2014kwa,Ding:2015uha,Piorkowska-Kurpas:2020rfy}. In the following, we will show that the microlensing case is relevant for a broad range of lens and source masses.

\subsection{Point mass}
\label{sec:pointmass}

In order to get a rough estimate of the lens and source masses for which the interference effect is detectable, we start with the simplest case where the lens can be approximated as a point mass. In this case $M_{\rm cyl}(x) = M_l$ is a constant, and, using Eq.~\eqref{eq:psix}, we get $\psi(x) = \ln x$ by choosing $\xi_0$ to be the Einstein radius, 
\be
    \xi_0 = \sqrt{\frac{4 M_l D_l D_{ls}}{D_s}} \equiv R_E(M_l) \,.
\ee 
The dimensionless frequency is $w = 8\pi M_{lz} f$, where $M_{lz} \equiv (1+z_l) M_l$. For a point mass lens the amplification factor $F(f)$ depends only on two parameters: $y$ and $M_{lz}$.

%\begin{figure}
%    \centering
%    \includegraphics[width=0.82\textwidth]{sinwT.pdf} 
%    \caption{The color coding illustrates the oscillations of the integrand of the point mass lens amplification function in the lens plane for $y=1$, $f=50\,$Hz, $z_l = 0.23$ and for two different lens masses. The lens lies at the origin, the black arrow shows the projected source position, and the positions of the minimum and the saddle point of the time delay function are shown by the red arrows.}
%    \label{fig:oscillations}
%\end{figure}

The point mass lens potential supports one path that minimizes and one that is a saddle point of the time delay function. These are at $\vect{x}_\pm(\vect{y}) = \pm x_\pm(y) \vect{y}/y$, where $x_\pm(y) = (\pm y+\sqrt{y^2+4})/2$. By taking $\phi(y) = (x_+(y)-y)^2/2-\ln{x_+(y)}$, the time delay function for the former path is zero. The integral over $\vect{x}$ in $F$ can be done in terms of the gamma function and the confluent hypergeometric function of the first kind as~\cite{Takahashi:2003ix}
\be \label{eq:FPM}
    F(w,y) = \exp\left[\frac{\pi w}{4}+i\frac{w}{2}\left(\ln{\frac{w}{2}-2\phi(y)}\right)\right] \Gamma\left(1-\frac{iw}{2}\right) {}_1 F_{1}\left(i\frac{w}{2};1;\frac{i}{2}w y^2\right) \,.
\ee
The magnifications for the geometric optics paths crossing the lens plane at $\vect{x} = \vect{x}_\pm$ are $\mu_{\pm}(y) = 1/2 \pm (2+y^2)/(2y\sqrt{y^2+4})$, and their dimensionless arrival time difference is
\bea \label{eq:time_dif}
    \Delta T(y) = \frac{y\sqrt{y^2+4}}{2}+\ln{\left(\frac{\sqrt{y^2+4}+y}{\sqrt{y^2+4}-y}\right)} \,.
\eea
The geometric optics approximation works for $w \gg 1/\Delta T(y)$. %As shown in Fig.~\ref{fig:oscillations}, in this limit the integrand $e^{i\omega T(\vect{x},\vect{y})}$ oscillates very rapidly except around the paths that extremise the time delay function. This makes the contributions to the integral of paths outside the extremal ones negligible. 
The time delay difference between the geometric optics paths is $\Delta t \equiv 4 M_{lz} \Delta T$.

\begin{figure}
    \centering
    \includegraphics[width=\textwidth]{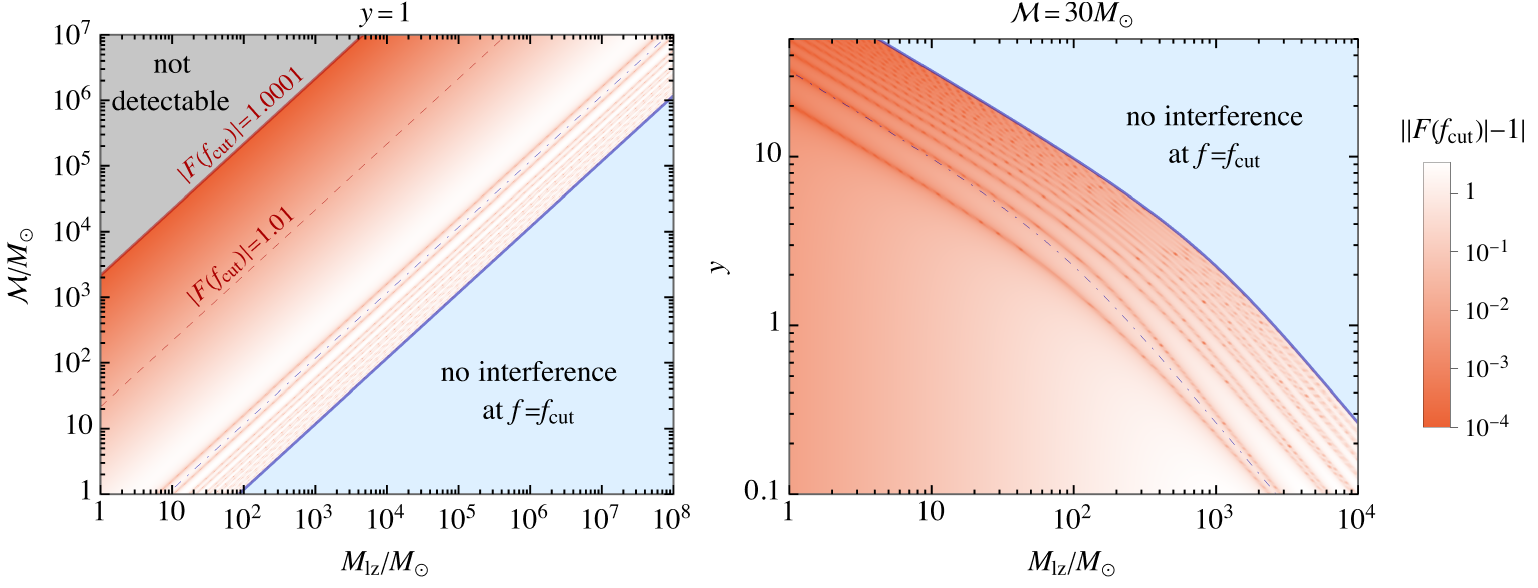}
    \caption{In the light blue region the left-hand side of the interference condition~\eqref{eq:interferencePM} is larger than one, and the dotdashed contour indicates where it is 0.1. In the light gray region $|F(f_{\rm cut})|<1.0001$. The symmetric mass ratio of the source binary is fixed to $\eta = 0.23$.}
    \label{fig:consistency}
\end{figure}

Let us check when the interference condition~\eqref{eq:interference} is fulfilled. During the inspiral phase of a compact binary coalescence, the frequency of the GW signal evolves as
\be
    \frac{\td f}{\td t} = \frac{3 f}{8 (t_c-t)} \,,
\ee 
where $t_c$ denotes the time of coalescence of the binary. The frequency changes most rapidly at the end of the inspiral phase which we approximate by the innermost stable circular orbit (ISCO),
\be
    f_{\rm cut} = \frac{\eta^{3/5}}{6^{3/2}\pi \mathcal{M}} \,,
\ee
where $\eta$ denotes the symmetric mass ratio of the binary. Thus, at the ISCO frequency, the interference condition~\eqref{eq:interference} for a point mass lens is given by
\be \label{eq:interferencePM}
    \frac{8}{135} \Delta T \,\eta^{8/5} \,\frac{M_{lz}}{\mathcal{M}} \ll 1 \,.
\ee
In the light blue region in Fig.~\ref{fig:consistency} the left-hand side of Eq.~\eqref{eq:interferencePM} is larger than one, and therefore the interference assumption is violated at the end of the inspiral. As shown in the right panel of Fig.~\ref{fig:consistency} , for $y = \mathcal{O}(1)$ and the sources similar to those that LIGO-Virgo has observed, the interference condition~\eqref{eq:interferencePM} is fulfilled as long as $M_{lz} \lsim 10^3 M_{\odot}$. For heavier sources, the interference condition holds up to higher lens masses, as shown in the left panel.

For the lensing to be potentially detectable the mass of the lens cannot be several orders of magnitude lighter than the binary as in that case the amplification factor would not significantly deviate from unity. In Fig.~\ref{fig:consistency} we show by the color coding $||F(f_{\rm cut})|-1|$ in logarithmic scale. Along the red dashed contour in the left panel this quantity equals $10^{-2}$ and in the gray region it is smaller than $10^{-4}$. The further we go toward the upper left corner of that plot, the more challenging the detection of the microlensing effect becomes. The boundaries of both "no interference" and "not detectable" regions scale as $\mathcal{M} \propto M_{lz}$, and there is a broad range of lens masses that can potentially be probed through microlensing of GW signals from binaries. Notice also, that at times much before the binary reaches the ISCO, the interference condition is satisfied even in the light blue region but if the merger is outside of the sensitivity range of the detector, then also the region that is not detectable moves to higher lens masses.

\subsection{Uniform density sphere}

A uniform sphere is characterized by its mass $M_l$ and radius $R$, and the mass density of the lens is
\be
\rho(r) = \frac{3M_l}{4\pi R^3} \,\theta(R-r) \,,
\ee
where $\theta$ denotes the Heaviside step function. The projected mass within radius $r$ in the lens plane is
\be
M_{\rm cyl}(r) = M_l \left[ 1 - \left(1-\frac{r^2}{R^2}\right)^{\!3/2} \theta(R-r) \right] \,.
\ee
Using Eq.~\eqref{eq:psix}, choosing $\xi_0$ in the same way as for a point mass lens and defining $b\equiv R/\xi_0$, we find that the deflection potential for a uniform density sphere is
\be \label{eq:psiStar}
\psi(x) = 
\begin{cases}
\frac13 \sqrt{1 - \frac{x^2}{b^2}} \left( \frac{x^2}{b^2}- 4 \right) + \ln\!\left(b + \sqrt{b^2-x^2} \right)  \,, & x\leq b \,, \\
\ln x \,, & x>b \,.
\end{cases}
\ee
As expected, the deflection potential coincides with that of the point mass lens for $x>b$. The dimensionless frequency is the same as in the point mass case, $w = 8\pi M_{lz} f$. The amplification function can be computed utilizing the analytical result of the point mass approximation~\eqref{eq:FPM}, so that the integral over the lens plane needs to be performed numerically only up to $x=b$.\footnote{Notice that $\phi(y)$ still needs to be calculated with the full lens potential~\eqref{eq:psiStar}. Given $y$ and $b$, we resolve $\phi(y)$ numerically.}

\begin{figure}
    \centering
    \includegraphics[width=\textwidth]{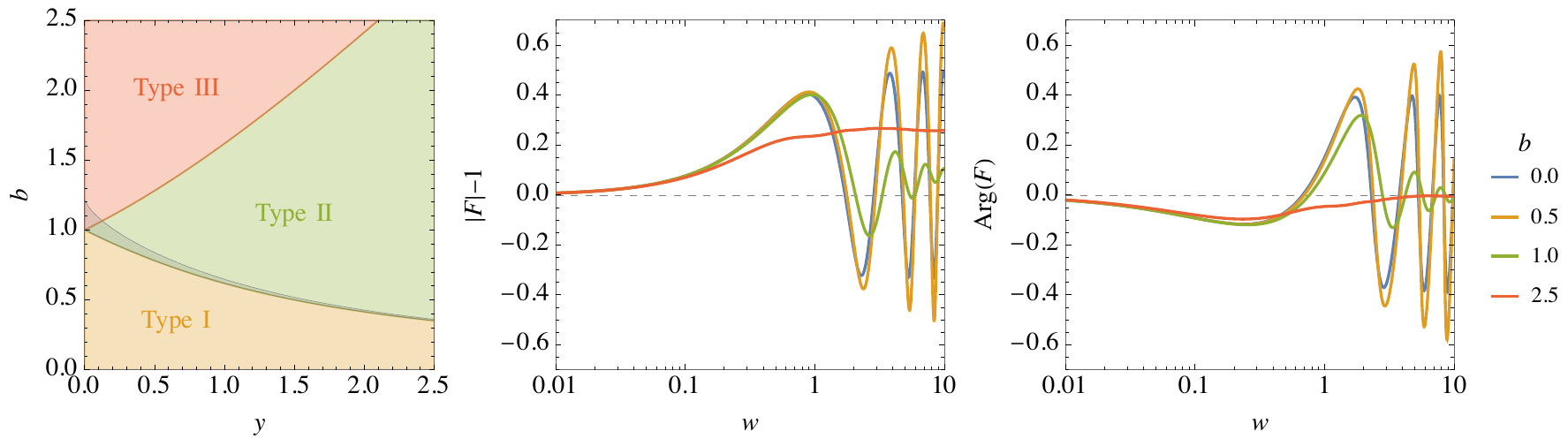}
    \caption{The left panel shows the regions corresponding to the three types of lensing by a uniform sphere discussed in the main text. The middle and right panels show the modulus and argument of the amplification function $F$ as a function of the dimensionless frequency $w$ for different lens sizes $b$ and $y=1$. The $b=0$ case corresponds to the point mass lens, $b=0.5$ is what we refer to as type~I, $b=1$ as type~II, and $b=2.5$ as type~III lensing, as shown in the left panel.}
    \label{fig:AmpDMS}
\end{figure}

There can be at most two extremal paths that cross the lens plane outside of the sphere ($x>b$). These are the same paths as for the point mass lens. The amplification factor for the sphere is heavily influenced by whether these paths exist. Also extremal paths that pass through the interior ($x\leq b$) of the sphere can exist. These can be easily found numerically. We find that the following three cases can be realized:
\begin{itemize}
\item Type I: if 
\be \label{ycrit_sol}
   0 < b\leq 1 \quad {\rm and} \quad y \leq \frac{1-b^2}{b}
\ee
there are two extremal paths at $x>b$ and one at $x\leq b$.
\item Type II: if
\be
b > 0 \quad {\rm and} \quad y>\frac{|1-b^2|}{b}
\ee 
there is one extremal path at $x>b$. In a narrow range just above the lower bound of $y$, there is also two extremal paths at $x<b$.
\item Type III: if 
\be 
b>1 \quad {\rm and} \quad y \leq \frac{b^2-1}{b} \,.
\ee 
and there is one extremal path at  $x\leq b$. In a small region at $b<\sqrt{3/2}$ two additional extremal paths at exist at $x<b$.
\end{itemize} 
In Fig.~\ref{fig:AmpDMS} we illustrate these cases: For the most compact objects type I is realized and the amplification function is very close to that of a point mass. The path crossing through the interior of the sphere may, however, boost the amplification, as can be seen by comparing the blue and yellow curves. Instead, for type II, shown by the green curves, the interference effect is smaller than for a point mass lens and the amplification converges to a constant deep in the geometric optics limit since the lens potential supports only one extremal time path. Finally, in the type III case, shown by the red curves, the amplification function does not have any significant oscillations, and the lens causes only a small overall amplification of the signal. However, even in this case the small increase in the amplification as a function of the frequency may give a chance to detect the lens. In the narrow band shown in gray shading in the left panel two additional extremal paths exist that cross through the sphere.

\subsection{Navarro–Frenk–White halo}
\label{sec:NFWlensing}

The Navarro–Frenk–White (NFW) density profile is characterized by the scale radius $r_s$ which separates the central $\rho \propto 1/r$ scaling from the outer $\rho \propto 1/r^3$:
\be
    \rho(r)=\frac{r_s^3 \rho_s}{r\left(r_s+r\right)^2} \,.
\ee
The projected mass within radius $r$ in the lens plane for the NFW profile is
\be
    M_{\rm cyl} = 4\pi \rho_s r_s^3\left[ \ln{\left(\frac{r}{r_s}\right)} + \frac{r_s}{\sqrt{r_s^2-r^2}} \arcoth  \left(\frac{r_s}{\sqrt{r_s^2-r^2}}\right) \right] \,, 
\ee
and, using Eq.~\eqref{eq:psix}, we find that the NFW deflection potential is given by
\be
    \psi(x) = \kappa \left[\ln ^2\!\left(\frac{x}{2 b}\right)-\artanh^2 \sqrt{1-\frac{x^2}{b^2}}\right] \,,
\ee
where we have defined dimensionless parameters $\kappa \equiv 2 \pi \rho_s r_s^3/M_{\rm v}$ and $b \equiv r_s/\xi_0$. The distances are scaled by $\xi_0 = R_E(M_v)$ and with this choice of $\xi_0$ the dimensionless frequency is $w = 8\pi (1+z_l) M_{\rm v} f$. For example, for a halo of mass $M_{\rm v} = 10^5 M_\odot$ at luminosity distance $D_{\rm sL} = 1$\,Gpc and a lens at luminosity distance $D_{\rm lL} = 0.5$\,Gpc, we get $\xi_0 \approx 2$\,pc.

The NFW parameters $r_s$ and $\rho_s$ are related, and can be expressed as functions of the halo viral mass $M_{\rm v}$, $r_s = r_s(M_{\rm v})$ and $\rho_s = \rho_s(M_{\rm v})$. Consequently, the prefactor $\kappa$ is fixed by $M_{\rm v}$. The parameter $b$ instead depends on $M_{\rm v}$ and the distances to the source and to the lens. 

\begin{figure}
    \centering
    \includegraphics[width=\textwidth]{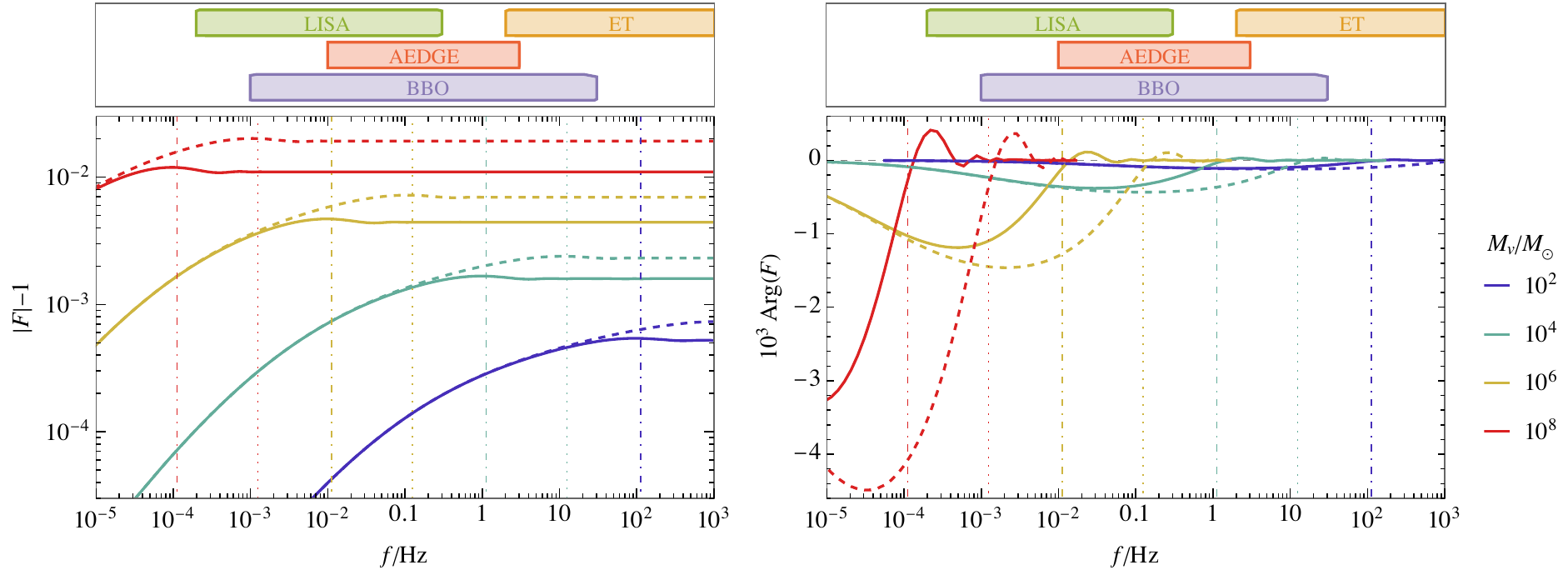}
    \caption{The modulus and the argument of the amplification function $F$ for NFW halo lens as a function of frequency for different virial masses of the halo. The solid and dashed curves correspond to $y=1$ and $y=0.3$, respectively, and the source and lens luminosity distances are fixed to $D_{\rm sL} = 5$\,Gpc and $D_{\rm lL} = 2.5$\,Gpc. The vertical lines indicate the position of the maximum of $|F|$. The frequency ranges relevant for different experiments are also indicated.}
    \label{fig:amp_nfw}
\end{figure}

We calculate the amplification function for the NFW lens using the numerical strategy described in~\cite{Takahashi:2004phd}. The modulus and the argument of the amplification function are shown for different halo viral masses in Fig.~\ref{fig:amp_nfw}. For $M_{\rm v} < 10^{10} M_\odot$ we find that $\kappa \lsim 0.3$ and $b > 2$. For these values of $\kappa$ and $b$ there is only one path that extremizes the time delay $T$. This path crosses the lens plane at $\vect{x} \approx \vect{y}$. As shown in Fig.~\ref{fig:amp_nfw}, the amplification function is similar to the Type III case for a uniform-density sphere lens, showing no clear interference pattern at the detector. 

In the geometric optics limit, the effect of the halo is a constant magnification of the signal that biases the apparent distance of the source. The magnification of the signal can be approximated by finding numerically the path that minimizes the time delay function using Eq.~\eqref{eq:amplification}, and the frequency above which the wave optics effects are no longer important is $w_{\rm geo} \approx 2/y^2$ which implies $f_{\rm geo} \approx 1/(4\pi y^2 (1+z_{\rm l}) M_{\rm v})$. This frequency gives approximately the position of the maximum in $|F|$, as indicated by the vertical lines in Fig.~\ref{fig:amp_nfw}.

For $M_{\rm v}>10^4 M_{\odot}$ the wave optics regime is at $f\lsim 1.6\,{\rm Hz}/(y^2(1+z_l))$. Therefore, LIGO and ET, which are sensitive at $f>1$\,Hz, can not distinguish the lensed signal from the unlensed one if the virial mass of the lens halo is $M_{\rm v} < 10^4 M_{\odot}$, unless $y\ll 1$ which makes the probability of such events very low. Instead, for $M_{\rm v} < 10^4 M_{\odot}$ the magnification that the NFW halos can generate is for realistic distances and impact parameters, at most $\mathcal{O}(10^{-3})$. This effect is so small that neither LIGO nor ET can distinguish the lensed signal from an unlensed one. Therefore, in the case of NFW halos, we will focus on LISA, AEDGE and BBO, which are sensitive at frequencies $f<1$\,Hz, and can probe the wave optics effects caused by halos with $M_{\rm v} > 10^4 M_\odot$.

High-definition galaxy simulations tend to prefer the Einasto profile, especially for low-mass DM halos~\cite{Wang:2019ftp}. The Einasto profile is
\be
\rho(r)=\rho_{-2}\exp\left[-\frac{2}{\alpha}\left(\left(\frac{r}{r_{-2}}\right)^\alpha-1\right)\right] \,,
\ee
where  $r_{-2}$ and $\rho_{-2}$ denote radius and density where the logarithm scale slope of the profile is $-2$. In the same way as for the NFW profile, these are functions of $M_{\rm v}$. For the shape parameter we consider the value $\alpha=0.158$. The relevant quantities for lensing can be analytically expressed in terms of Fox $H$ function~\cite{Retana-Montenegro:2012dbd}. We have, however, simply computed them numerically. %As shown in Fig.~\ref{fig:einasto} the difference in the deflection potentials for the NFW and Einasto profiles is very small.
We have found that the Einasto and the NFW profiles are sufficiently similar to not produce any significant differences in the scales relevant to GW lensing. Therefore, in the following we consider only the NFW profile for the CDM halos.

%\begin{figure}
%    \centering
%    \includegraphics[width=0.94\textwidth]{einasto.pdf}
%    \caption{The solid and dashed curves show the mass density profiles (left panel) and the deflection potentials (right panel) for NFW and Einasto halos, respectively, of virial mass $M_{\rm v}=10^5 M_{\odot}$ at $D_{\rm lL} = 0.5$\,Gpc and a source at $D_{\rm sL}= 1$\,Gpc. The vertical dashed lines in the left panel correspond to those in the right panel.}
%    \label{fig:einasto}
%\end{figure}

\section{Differentiating the lensed signal from unlensed one}
\label{sec:analysis}

Consider data that consists of a GW signal $\phi$ and noise that we characterize by the noise power spectral density $S_{\rm n}(f)$.\footnote{In addition to the detector noise, we account for the noise coming from the unresolved white dwarf and black hole binary backgrounds~\cite{Lewicki:2021kmu}.} The standard matched filtering analysis searches for the parameters $\vect\theta$ of a given template $\phi_T$ that maximize the log-likelihood~\cite{Jaranowski:2005hz}
\be \label{eq:likelihood}
    \ln \Lambda_T(\phi) = (\phi|\phi_T(\vect\theta)) - \frac12 (\phi_T(\vect\theta)|\phi_T(\vect\theta)) \,,
\ee
where $(a|b)$ denotes the noise-weighted inner product,
\be \label{inner_product}
    (a|b) \equiv 2 \int \!\td f \,\frac{\tilde a^*(f) \tilde b(f) + \tilde a(f) \tilde b^*(f)}{S_{\rm n}(f)} \,.
\ee
For an optimal template $\phi_T = \phi_{\rm opt}$ we can find parameters for which $\phi = \phi_{\rm opt}$. If we know the optimal template, we can calculate how well another template $\phi_T$ fits the signal by maximizing the log-likelihood~\eqref{eq:likelihood} over the parameters of the template and compare it to the optimal template by calculating the log-likelihood difference
\bea \label{eq:chis}
    \Delta\chi^2 &\equiv 2\left[\ln \Lambda_{\rm opt}(\phi) - \max_{\vect\theta} \ln \Lambda_T(\phi) \right] \\
    %&= \min_{\vect\theta}\left[ (\phi_T(\vect\theta)|\phi_T(\vect\theta)) - 2(\phi_{\rm opt}|\phi_T(\vect\theta)) + (\phi_{\rm opt}|\phi_{\rm opt})) \right] \\
&= \min_{\vect\theta} \,(\phi_{\rm opt}-\phi_T(\vect\theta)|\phi_{\rm opt}-\phi_T(\vect\theta)) \\
&= 4 \min_{\vect\theta} \int \!\td f \,\frac{|\tilde\phi_{\rm opt}(f)-\tilde\phi_T(f;\vect\theta)|^2}{S_{\rm n}(f)}\,.
\eea
Assuming Gaussian statistics, if $\Delta\chi^2 > 4.00,6.18,\,8.02,\,9.72$ for one, two, three or four parameter model, respectively, we can differentiate the templates $\phi_T$ and $\phi_{\rm opt}$ at $2\sigma$ confidence level. 

We use Eq.~\eqref{eq:chis} to quantify the detectability of the lens effect in the GW signal from a black hole binary.\footnote{The same method can be used also to study whether different types of lenses can be differentiated from each other. We leave that analysis for future studies.} In this case we take $\phi_{\rm opt}$ as the lensed signal with given lens and source parameters, and try to fit that with the unlensed template. The unlensed template is given by $\tilde \phi_T(f;\vect\theta) = A(f;\vect\theta) e^{-i\Psi(f;\vect\theta)}$. We consider the lowest order approximations for the amplitude $A(f;\vect\theta)$ and the phase $\Psi(f;\vect\theta)$:
\be \label{eq:unlensedtemplate}
    A(f;\vect\theta) = \sqrt{\frac{5}{24}} \frac{\mathcal{M}^{5/6}}{\pi^{2/3} D_L} f^{-\frac76} \,, \qquad \Psi(f;\vect\theta) = 2\pi ft_c - \phi_c + \frac{3}{128}\left(\pi\mathcal{M}f\right)^{-\frac{5}{3}} \,.
\ee
This template is defined by 4 parameters: $\vect\theta = \{ \mathcal{M}, D_L, t_c, \phi_c \}$. The luminosity distance is related to $D_s$ via $D_L = (1+z_s) D_s$, and the binary chirp mass is related to the binary total mass $M$ and the binary mass ratio $q$ by $\mathcal{M} = (1+z_s) M (q/(1+q))^{3/5}$.

The lensed GW signal is given by $\tilde\phi_{\rm opt}(f) = F(f;\vect\theta_{l,{\rm opt}}) A(f;\vect\theta_{\rm opt}) e^{-i\Psi(f;\vect\theta_{\rm opt})}$, where $F(f;\vect\theta_{l,{\rm opt}})$ is the lensing amplification factor defined by the lens parameters $\vect\theta_l = \vect\theta_{l,{\rm opt}}$. To quantify how accurately this signal can be fitted with the unlensed GW template, we evaluate Eq.~\eqref{eq:chis} finding the minimum over the 4 parameters of the unlensed GW template. We find in all cases that the minimum is reached for $\mathcal{M} \approx \mathcal{M}_{\rm opt}$, and for $t_c \approx t_{c,{\rm opt}}$. In practice we therefore scan only over $\lambda \equiv (D_L/D_{L,{\rm opt}})^{-1}$ and $\delta \phi_c \equiv \phi_c - \phi_{c,{\rm opt}}$. In this case, we can write $\Delta\chi^2$ as
\be\label{unlensed}
\Delta\chi^2 = 4 \min_{\lambda,\,\delta\phi_c} \int \!\td f \,\frac{A(f;\vect\theta_{\rm opt})^2 |F(f;\vect\theta_{l,{\rm opt}})-\lambda e^{-i \delta \phi_c}|^2}{S_{\rm n}(f)} \,.
\ee
Notice that if the merger does not happen within the sensitivity window of the observatory, then $\Delta\chi^2 \propto {\mathcal{M}^{5/3}}/{D_L^2}$.

%\begin{figure}
%    \centering
%    \includegraphics[width=0.56\textwidth]{Sn.pdf}
%    \caption{Noise power spectral density of LIGO at its design sensitivity and various proposed GW detectors. The contributions coming from the unresolved binary backgrounds are shown by the dashed curves.}
%    \label{fig:Sn}
%\end{figure}

We evaluate $\Delta \chi^2$ in the frequency range corresponding to at most a year of inspiralling observation finishing at the ISCO orbit. 
%The noise power spectral densities that we consider are shown in Fig.~\ref{fig:Sn}. We account also for the noise contribution coming from the unresolved white dwarf and black hole binary backgrounds~\cite{Lewicki:2021kmu}, also shown in Fig.~\ref{fig:Sn}.
Once $\Delta \chi^2$ is computed as a function of the impact parameter $y$, it is possible to invert the relation to find the maximum impact parameter $y_{\rm max}$ that leads to a detectable lens effect. We define $y_{\rm max}$ so that for $y<y_{\rm max}$ the lensed template can be separated from the unlensed one by more than $2\sigma$ confidence level. As the unlensed template, Eq.~\eqref{eq:unlensedtemplate}, includes 4 parameters, this means that $\Delta \chi^2 > 9.72$ for $y<y_{\rm max}$. The maximal detectable impact parameter is relevant as it determines the detectable lens cross-section and consequently the probability that the lens effect can be detected in a given GW signal. We return to this in Sec.~\ref{sec:probability}. In the following, we compute $y_{\rm max}$ for different lenses, sources, and detectors.

\subsection{Cold dark matter halos}

For CDM halos, we consider the NFW density profile described in Sec.~\ref{sec:NFWlensing}. As discussed in the Introduction, the most interesting mass range is $M_{\rm v} \lsim 10^8M_\odot$ as these halos don't include many stars and their detection is challenging by conventional techniques. As shown in Sec.~\ref{sec:NFWlensing}, this in fact is the mass range where the effect of lensing of GW signals by DM halos is within the frequency window of the future GW interferometers. 

\begin{figure}
    \centering
    \includegraphics[width=1\textwidth]{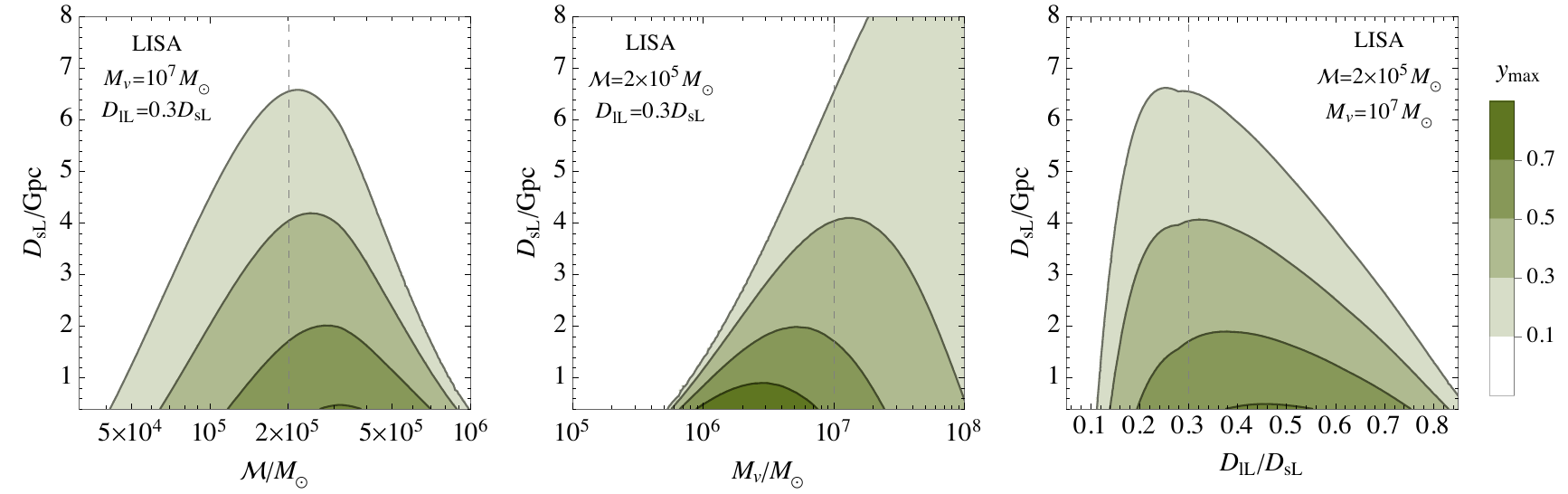} \\
    \includegraphics[width=1\textwidth]{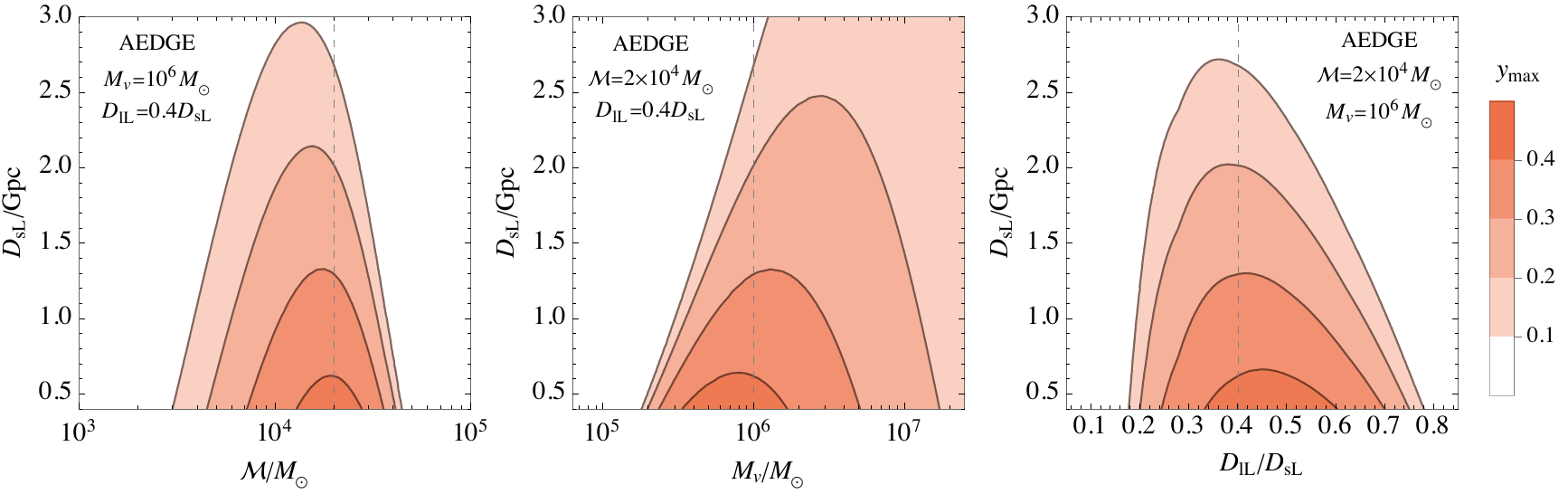} \\
    \includegraphics[width=1\textwidth]{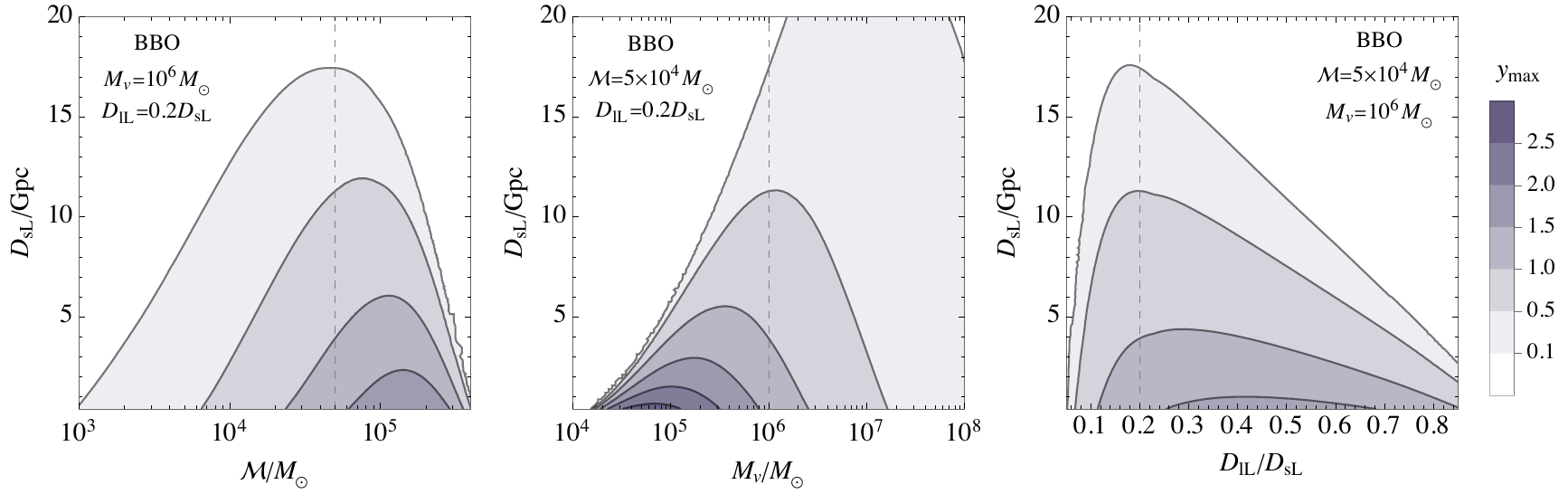}
    \caption{The maximum value of the impact parameter $y_{\rm max}$ that allows for detection of the lens effect by an NFW halo as a function of the binary chirp mass $\mathcal{M}$, the lens halo virial mass $M_{\rm v}$, and the source and lens luminosity distances $D_{\rm sL}$ and $D_{\rm lL}$. The symmetric mass ratio of the binary is fixed to $\eta = 0.23$.}
    \label{fig:ymaxNFW}
\end{figure}

In Sec.~\ref{sec:NFWlensing}, we show why LIGO and ET cannot probe the NFW halos through gravitational lensing. Instead, as LISA, AEDGE and BBO are sensitive at lower frequencies, they can probe the lens effects from DM halos of virial mass $M_{\rm v} > 10^4 M_\odot$ for which the lens effect can be relatively strong, about a $1\%$ effect over the unlensed signal. We have computed $\Delta \chi^2(y)$ for differentiating the lens effect from the unlensed template for these experiments, and the corresponding maximal impact parameter $y_{\rm max}$. As shown in Fig.~\ref{fig:ymaxNFW}, each of these observatories can detect the lens effect for sources up to $\mathcal{O}({\rm Gpc})$ distances with a reasonably large $y_{\rm max}$. The mass ranges in both the chirp mass and the virial mass reflect the frequency ranges of these experiments. The lens effect becomes easier to detect the closer the source is, as for such sources the signal is stronger. The Einstein radius of the system vanishes at the lens distances $D_{\rm lL} = 0$ and at $D_{\rm lL} = D_{\rm sL}$, and consequently such cases are not detectable.

\subsection{Fuzzy dark matter halos}

\begin{figure}
    \centering
    \includegraphics[width=0.95\textwidth]{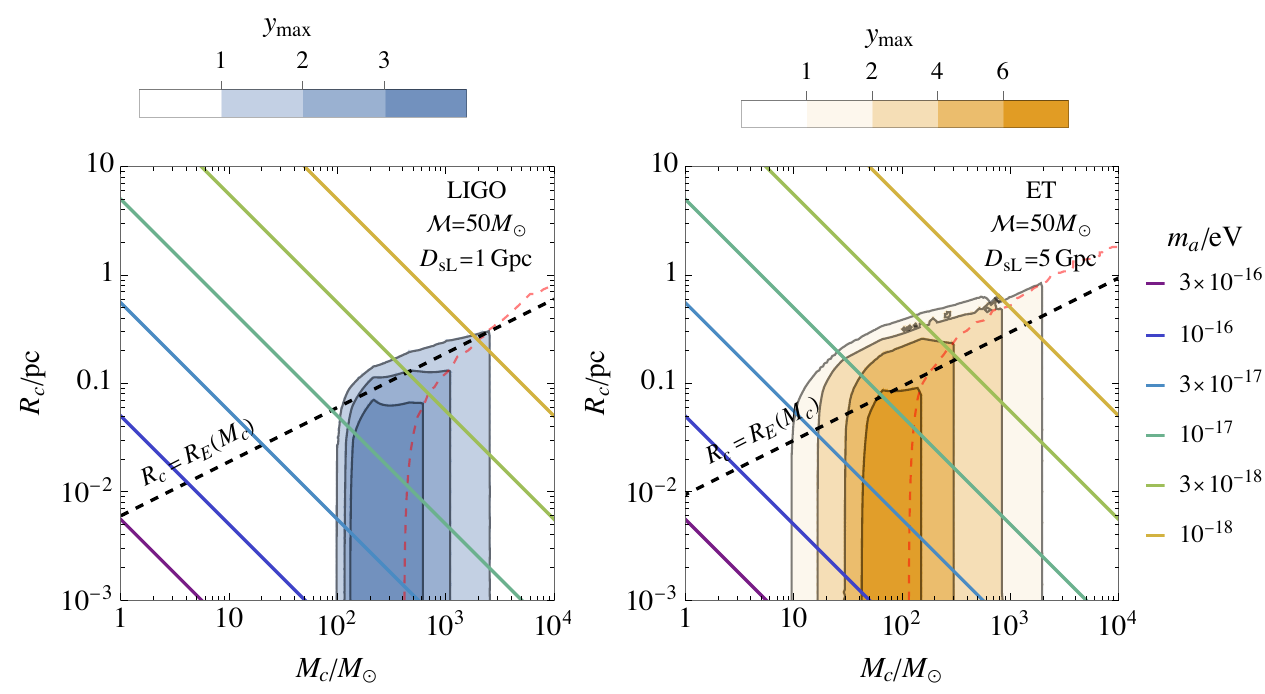}
    \caption{The color shading shows the maximum value of the impact parameter $y_{\rm max}$ that allows for detection of the lens effect by a uniform density sphere of mass $M_c$ and radius $R_c$. The binary chirp mass and luminosity distance are fixed as shown in the plot, the symmetric mass ratio of the binary is fixed to $\eta = 0.23$ and the lens distance to $D_{\rm lL} = 0.5 D_{\rm sL}$. The lines show the $M_c-R_c$ relation for DM core for different axion masses $m_a$, following Eq.~\eqref{eq:Rc}. Right from the red dashed curve $y_{\rm max}$ is set by the interference condition, and the black dashed curve shows the point mass Einstein radius, $R_c=R_E(M)$.}
    \label{fig:ymaxMR}
\end{figure}

DM halos consisting of fuzzy DM are expected to have relatively compact cores. We approximate them with the uniform density sphere deflection potential. In Fig.~\ref{fig:ymaxMR} we show in the mass-radius plane of the uniform density spheres the maximum impact parameter $y_{\rm max}$ that is detectable with LIGO and ET. If the radius $R_c$ of the object is much smaller than the Einstein radius of a point mass lens of the same mass $M_c$, $R_c\ll R_E(M)$, then the object behaves as a point mass and $y_{\rm max}$ does not depend on $R_c$. For $R_c\gsim R_E(M)$ the lens effect becomes suppressed, but as shown in Fig.~\ref{fig:ymaxMR}, ET can detect objects whose radius is almost an order of magnitude larger than $R_E(M)$. 

The properties of the cores have been studied by numerical simulations of structure formation of fuzzy DM~\cite{Schive:2014dra,May:2021wwp,Chan:2021bja}. These simulations give us predictions for the mass-radius relationships of cores in these halos as well as the core mass/halo mass ratio. There is some uncertainty with regards to the expected ratio between the core and halo mass but the predictions for the mass-radius relations of the cores themselves are more settled and neglecting the scatter found in these mass-radius relationships, we use the fit provided in~\cite{Schive:2014dra} which gives
\be \label{eq:Rc}
    R_c = 5.0\,{\rm pc} \, \left(\frac{m_a}{10^{-17}\,{\rm eV}}\right)^{\!-2} \left(\frac{M_c}{M_\odot}\right)^{\!-1} ,
\ee
where $m_a$ denotes the fuzzy DM particle mass. We overlap in Fig.~\ref{fig:ymaxMR} this mass-radius relationship for different fuzzy DM masses. We can see that LIGO is sensitive, assuming some realistic impact parameters, to DM masses from $10^{-16}$ to $3\times 10^{-18}$\,eV. ET can instead probe a wider range of masses going down to $10^{-18}$\,eV and lighter less compact cores.

\begin{figure}
    \centering
    \includegraphics[width=0.8\textwidth]{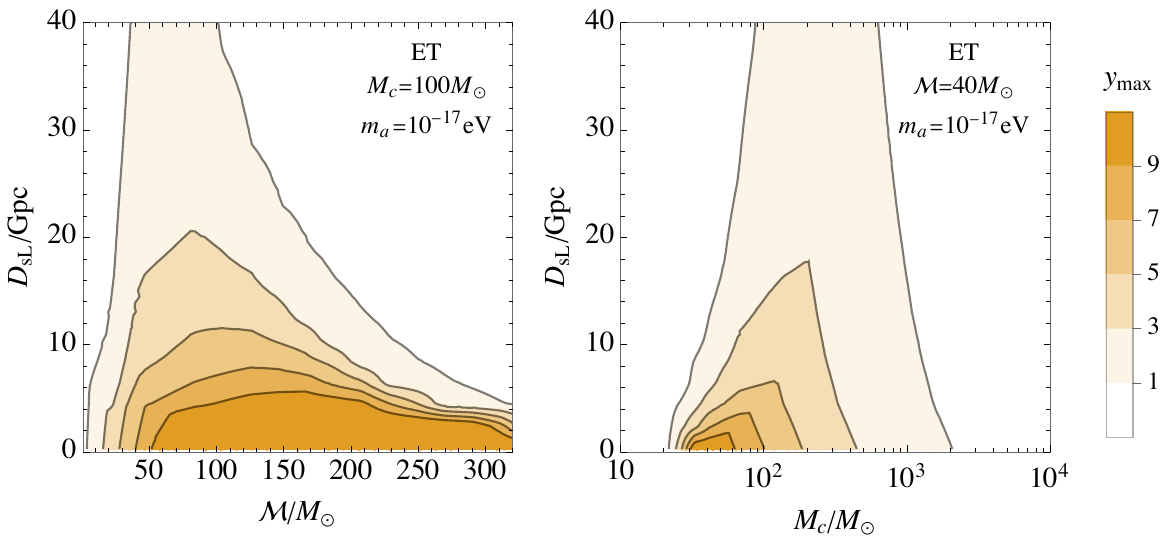}
    \caption{Maximum value of the impact parameter $y_{\rm max}$ for which ET can detect microlensing by a fuzzy DM core for the DM mass $m_{a}=10^{-17}\rm eV$ as a function of the luminosity distance and the binary chirp mass (left panel) or the core mass (right panel).}
    \label{fig:ymaxfuzzy}
\end{figure}

In Fig.~\ref{fig:ymaxfuzzy} we show the maximum impact parameter as a function of the source parameters and the core mass for a fixed DM mass $m_a = 10^{-17}$\,eV. The maximum impact parameter monotonically decreases with the distance and has a maximum in the chirp mass at around $100 M_{\odot}$. This is expected because if the binary is heavier the GW frequency starts to be too low to be detected with ET. The maximum impact parameter grows up to $M_c\gsim 100 M_\odot$ where the paths take too much time two arrive at the detector and the interaction is no longer satisfying the interference condition. 

\section{Prospects for gravitational wave experiments}
\label{sec:probability}

The probability for an event to be lensed assuming a Poisson distribution is
\be \label{eq:Pl}
    P_l = 1-e^{-\tau} \,,
\ee 
where $\tau$ denotes the optical depth associated with the lensing cross-section. The optical depth is given by 
\be
    \tau(z_s) = \int_0^{D_s} \td V \, n(D_l) \frac{\sigma(D_l)}{4\pi D_l^2} \,= \int_0^{z_s} \td z_l \frac{n(z_l) \sigma(z_l)}{(1+z_l) H(z_l)} \,,
\ee
where $\td V = 4\pi D_l^2 \td D_l$ denotes the differential volume element, $D_l$ is the angular diameter distance of the lens and $z_l$ the corresponding redshift, $n$ is the number density of the lenses and $H$ the Hubble expansion rate. The cross-section for detecting the lens effect in a given GW signal is
\be
    \sigma = \pi \xi_{\rm max}^2 = \pi \xi_0^2 y_{\rm max}^2 \,,
\ee
where $\xi_{\rm max} = \xi_0 y_{\rm max}$ is the maximal impact parameter of the source-lens-detector system for which the lens effect is detectable.

Accounting for the present-day halo mass function, $\td n_0/\td\!\log_{10}\!M_{\rm v}$, and assuming that the cross-section depends directly on the virial mass of the halo, we get
\be \label{eq_tau1}
    \tau(z_s) = \int_0^{z_s} \td z_l \int \td\!\log_{10}\!M_{\rm v} \, \frac{\td n}{\td\!\log_{10}\!M_{\rm v}} \frac{\sigma(z_l,M_{\rm v})}{(1+z_l) H(z_l)} \,.
\ee
In the case of lensing by the cores of fuzzy DM halos, we also need to account for the spread in the halo mass - core mass relationship~\cite{Chan:2021bja}, and the optical depth is given by
\be
    \tau(z_s) = \int_0^{z_s} \td z_l \int \td\!\log_{10}\!M_{\rm v} \, \frac{\td n}{\td\!\log_{10}\!M_{\rm v}} \int \td\!\log_{10}\!M_{\rm c}\, f(M_{\rm c},M_{\rm v}) \frac{\sigma(z_l,M_{\rm c})}{(1+z_l) H(z_l)} \,,
\ee
where $f(M_{\rm c},M_{\rm v})$ denotes the probability distribution function of the core mass $M_{\rm c}$ for a halo of virial mass $M_{\rm v}$, which is normalized such that $\int \td\!\log_{10}\!M_{\rm c}\, f(M_{\rm c},M_{\rm v}) = 1$.

From the lensing probability and the rate of GW events, we can calculate the expected number of events that a given detector will see:
\be \label{eq:Nl}
    N_l = \mathcal{T} \int \td \lambda \, p_{\rm det} \, P_l \, ,
\ee
where $\mathcal{T}$ denotes the observation period and $p_{\rm det}$ is the detection probability that accounts for averaging over the binary sky location, binary inclination, and polarization of the signal~\citep{Gerosa:2019dbe}. The differential merger rate per unit time is given by
\be
    \td \lambda = \frac{1}{1+z_s}\frac{\td R}{\td m_1 \td m_2} \frac{\td V_c}{\td z_s} \td m_1 \td m_2 \td z_s \,,
\ee
where $V_c$ is the comoving volume. Without the $P_l$ factor Eq.~\eqref{eq:Nl} gives the expected total number of detectable GW events.

\subsection{Cold dark matter halos}

In the case of CDM halos, which we describe with the NFW profiles we adopt the EPS formalism~\citep{Press:1973iz,Bond:1990iw} to calculate the galactic halo mass function,
\be\label{eq:cdmhmf}
	\frac{\td n(M,t)}{\td \ln M} = \frac{\rho_0}{M} \sqrt{\frac{2}{\pi}} \frac{\td \ln \sigma}{\td \ln M} \frac{\delta_c(z)}{\sigma (M)}e^{-\frac{\delta_c(z)^2}{2\sigma^2(M)}} \, ,
\ee
where $\rho_0$ is the background matter density today, $\sigma^2(M)$ is the variance of the matter fluctuations, $\delta_c(z)$ is given by the linear growth function $D(z)$ as $\delta_c(z) \approx 1.686/D(z)$~\citep{Dodelson:2003ft}. Using the results for $y_{\rm max}$ shown in Fig.~\ref{fig:ymaxNFW} we calculate the probability $P_l$ for a given GW source and detector by Eqs.~\eqref{eq:Pl} and ~\eqref{eq_tau1}.

\begin{figure}
    \centering
    \includegraphics[width=0.98\textwidth]{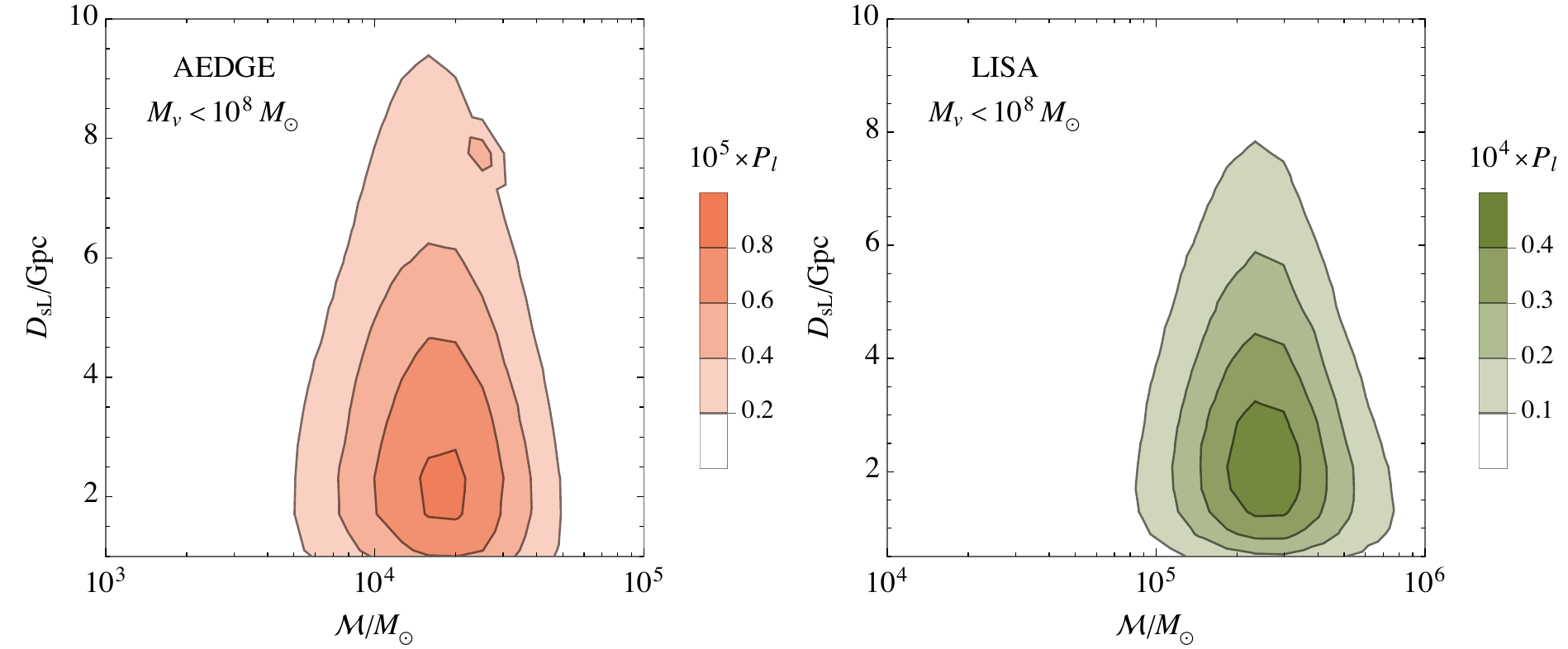} \\ \vspace{2mm}
    \includegraphics[width=0.92\textwidth]{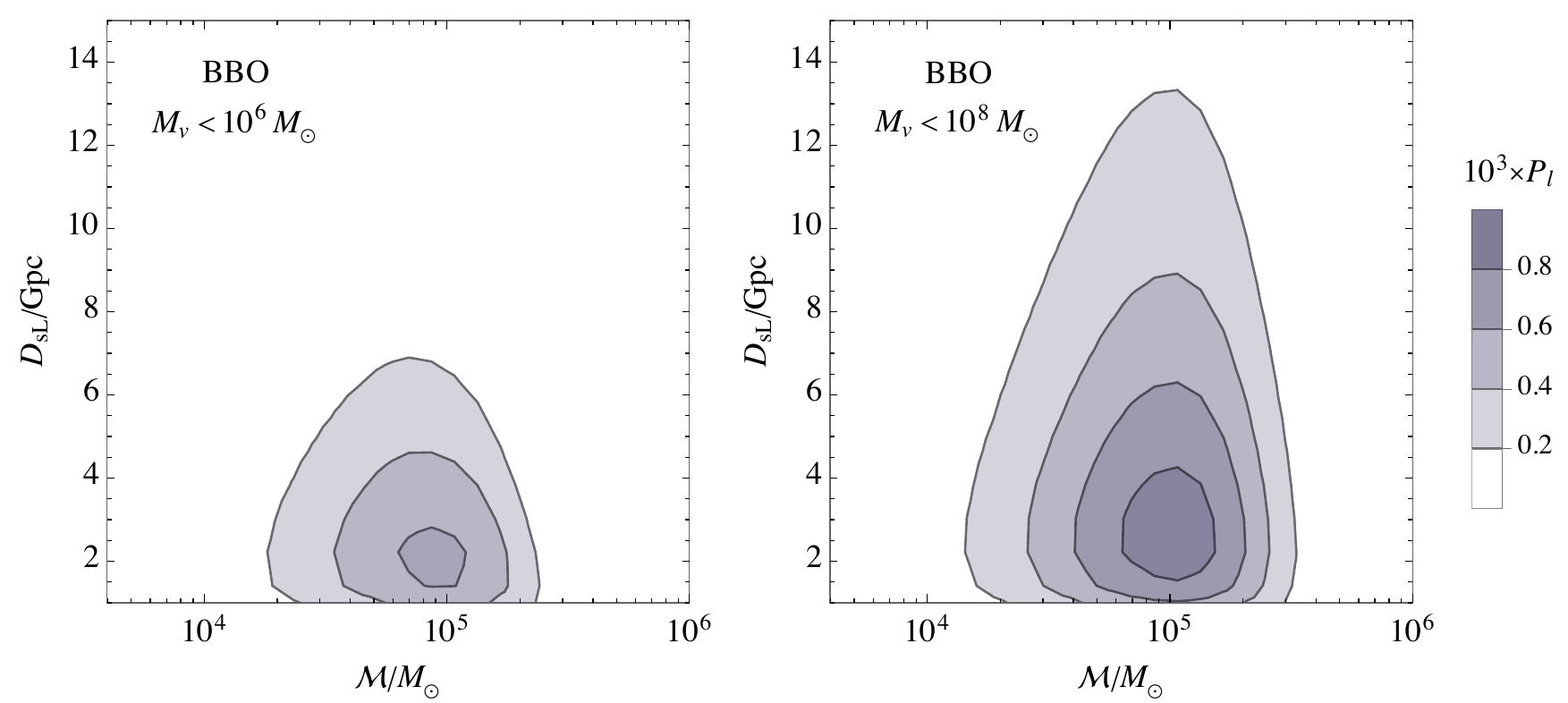}
    \caption{Probability $P_l$ that a GW signal emitted by a binary of chirp mass $\mathcal{M}$ and symmetric mass ratio $\eta = 0.23$ at distance $D_{\rm sL}$ is lensed by a NFW halo. The probabilities are integrated over the virial mass range shown in each panel with the CDM halo mass function.}
    \label{fig:Pnfw}
\end{figure}

The probability of an event to be lensed and detected by LISA, AEDGE and BBO is shown in Fig.~\ref{fig:Pnfw}. Although $y_{\rm max}$ increases with decreasing source distance as the signal becomes stronger, the maximal lensing probability is reached at a finite source distance because the number of lenses grows with volume. The maximal probability of detecting lensing by halos of mass $M_{\rm v} < 10^8 M_\odot$ is $P_l\approx 0.8\times 10^{-3}$ for a source of mass $\mathcal{M} \approx 10^5M_\odot$ at $D_{\rm sL} \approx 4$\,Gpc for BBO, $P_l \approx 5\times 10^{-5}$ for a source of mass $\mathcal{M} \approx 3\times 10^5M_\odot$ and $D_{\rm sL} \approx 3$\,Gpc for LISA, and $P_l \approx 10^{-5}$ for a source of mass $\mathcal{M} \approx 2\times 10^4M_\odot$ at $D_{\rm sL} \approx 3$\,Gpc for AEDGE. For BBO we also show the probability of lensing by halos smaller than $10^6 M_\odot$, which is at most $0.8\times 10^{-3}$.

The merger rate of intermediate-mass BHs (IMBHs) that we have considered as sources is uncertain. The observed dynamics of gas accretion and tidal disruption events provide possible candidates for nearby IMBHs~\cite{Whalen:2015hza,Lin_2018,Peng:2019ckh,Shen:2018rdx,Micic:2022xci,Toptun:2022dmr}, but the merger rate and population properties of IMBHs will be known only after enough binaries are detected with GW observatories. The component masses of the heaviest binary seen so far by LIGO-Virgo, GW190521, are $m_1 \approx 85M_\odot$ and $m_2 \approx 66 M_\odot$~\cite{LIGOScientific:2020iuh}. Searches have been performed for binaries of total mass up to $800 M_\odot$ and the non-observation of such binaries indicates an upper bound on their merger rate. For binaries consisting of two $100M_{\odot}$ BHs, the merger rate is smaller than $0.2\,{\rm Gpc}^{-3} {\rm yr}^{-1}$~\cite{LIGOScientific:2019ysc}. Merger rates have been proposed at the local universe  due to repeated merges in dense stellar environments like globular and nuclear clusters~\cite{Rasskazov:2019tgb,Fragione:2018vty,Fragione:2017blf,Sesana:2010wy,Fragione:2022avp}, but these are mostly mergers of IMBHs with stellar mass BHs. In this work, we have focused on the merger of BHs of a similar mass. Halo mergers may generate such IMBH binaries and consequently, their merger rate follows from the halo merger rate~\cite{Erickcek:2006xc, Barausse:2012fy, Valiante:2020zhj}.
%In Fig.~\ref{fig:nuclear_clusters} we show the expected merger rate for IMBHs predicted by halo merger theory in Ref.~\cite{Erickcek:2006xc}. Using this merger rate, 
Using the the expected merger rate for IMBHs predicted by halo merger theory in Ref.~\cite{Erickcek:2006xc}, we find that LISA is expected to see $3\times 10^{-7}$, AEDGE $1\times 10^{-5}$ and BBO $1\times 10^{-2}$ events per year that are lensed by halos of mass $M_{\rm v} < 10^8 M_\odot$. When integrating the probability over those halos that contain hardly any baryonic matter $\left(M_{\rm v}<10^6 M_{\odot}\right)$, the number of expected events drops to $1\times 10^{-3} $ events per year for BBO.

%\begin{figure}
%    \centering
%    \includegraphics[width=0.5\textwidth]{Nuclearcluster.pdf}
%    \caption{Merger rate estimate of binaries formed in halo mergers from~\cite{Erickcek:2006xc}. The curves show the merger rate of binaries where both of the components are in the indicated mass range.}
%    \label{fig:nuclear_clusters}
%\end{figure}

\subsection{Fuzzy dark matter halos}

\begin{figure}
     \centering
     \includegraphics[width=0.95\textwidth]{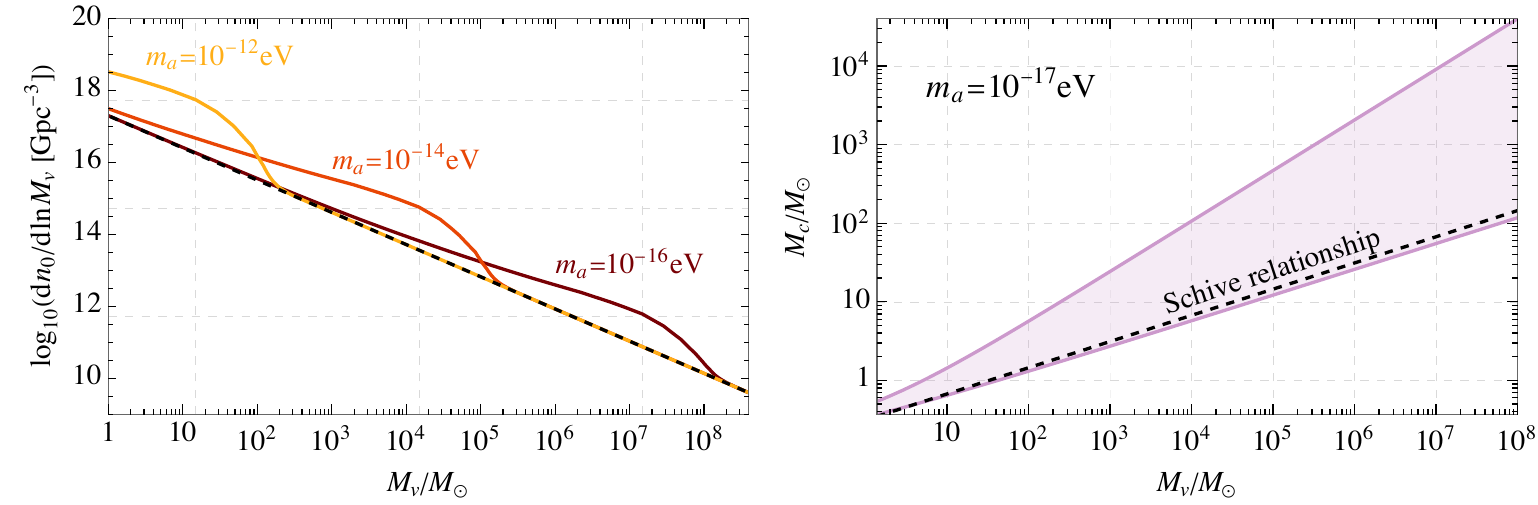}
     \caption{\emph{Left panel:} The present-day fuzzy DM halo mass function from~\cite{Fairbairn:2017sil} with a bump caused by miniclusters formed before matter-radiation equality. The dashed line shows the CDM halo mass function~\eqref{eq:cdmhmf}. \emph{Right panel:} The spread in the fuzzy DM core mass - halo mass relationship from~\cite{Chan:2021bja}.}
     \label{fig:hmf}
\end{figure}

In the fuzzy DM scenarios, the halo mass function may deviate significantly from the CDM case. First, the small scale fluctuations are suppressed which results in a low mass cut-off of the halo mass function at $M_v \sim (m_a/10^{-15}{\rm eV})^{-3/2} M_\odot$ (see e.g.~\cite{Kulkarni:2020pnb}). Second, in addition to the structures formed after matter-radiation equality, which we assume to follow the mass function~\eqref{eq:cdmhmf}, miniclusters that can have formed before matter-radiation equality contributes to the total halo mass function. This is relevant for the axionic fuzzy DM scenarios where the Peccei-Quinn symmetry breaks after the cosmic inflation~\cite{Kolb:1993zz}. Using the results presented in Ref.~\cite{Fairbairn:2017sil}, we find that the present day the DM halos generated from these miniclusters give rise to a bump in the halo mass function at $M_v \simeq 5\times 10^5 (m_a/10^{-15}{\rm eV})^{-3/2} M_\odot$ (more than 5 orders of magnitude above the cut-off mass) that exceeds the halo mass function~\eqref{eq:cdmhmf} by factor $\simeq 10 (m_a/10^{-15}{\rm eV})^{0.15}$. In the left panel of Fig.~\ref{fig:hmf}, we show this bump in the halo mass function for different values of the axion mass $m_a$. 

Moreover, the fuzzy DM halos have solitonic cores that we approximate with a sphere of constant density and radius given by Eq.~\eqref{eq:Rc}. The core mass $M_{\rm c}$ can have a range of values for a given halo mass $M_{\rm v}$. We estimate the scatter of the core masses using the results of Ref.~\cite{Chan:2021bja} assuming that the distribution of the core masses is uniform in logarithmic scale in the range $M_{{\rm c},1} < M_{\rm c} < M_{{\rm c},2}$ given by the fitting formula
\be
\frac{M_{{\rm c},j}}{M_\odot} = \beta_j \left( \frac{m_a}{10^{-17}{\rm eV}}\right)^{-3/2} + \gamma_j \left( \frac{m_a}{10^{-17}{\rm eV}}\right)^{3(\alpha_j-1)/2} \left( \frac{M_{\rm v}}{10^6 M_\odot}\right)^{\alpha_j}
\ee
with $\alpha_1 = 0.33$, $\beta_1 = 0.062$ and $\gamma_1 = 26$, and $\alpha_2 = 0.64$, $\beta_2 = 0.19$ and $\gamma_2 = 2.1\times 10^3$. We show this mass range for a benchmark value of $m_a$ in the right panel of Fig.~\ref{fig:hmf}. For comparison, the black dashed line shows the relationship obtained in Ref.~\cite{Schive:2014dra}, $M_{\rm c} = 31 (m_a/10^{-17}{\rm eV})^{-1} (M_{\rm v}/10^6 M_\odot)^{1/3}$.

\begin{figure}
     \centering
     \includegraphics[width=\textwidth]{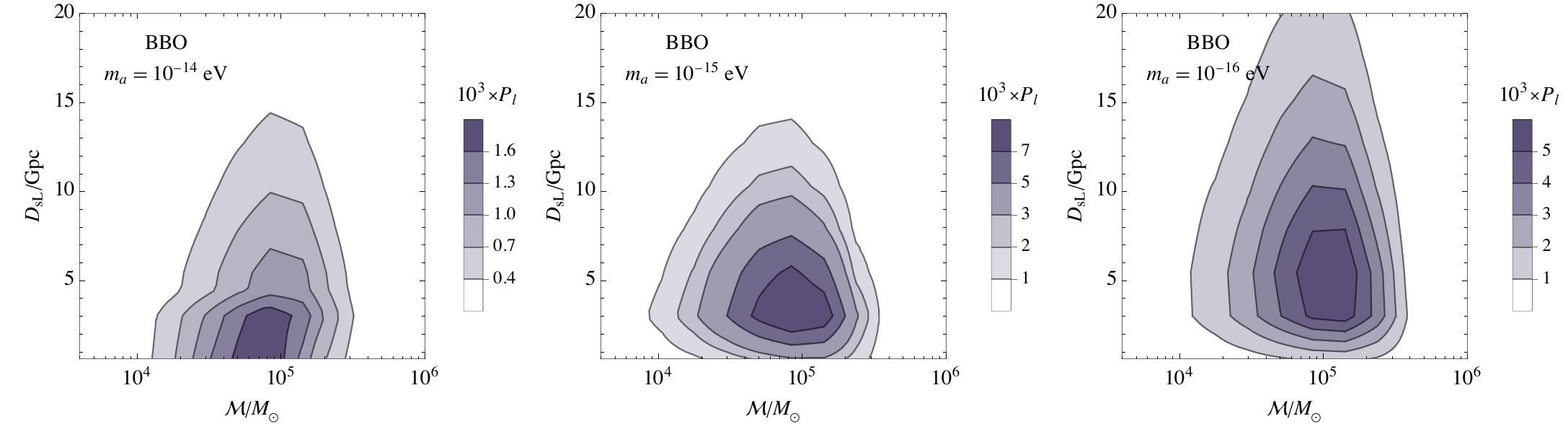}
     \caption{Probability of lensing by a NFW DM halo with $M_{\rm v} < 10^8 M_\odot$ considering the contribution of miniclusters for three fuzzy DM masses.}
     \label{fig:axion_bump}
\end{figure}

As discussed in Sec.~\ref{sec:pointmass}, the microlensing effect is negligible if the binary is much heavier than the lens. So, even if there are solitonic cores inside fuzzy DM halos, we will take into account only the lensing by the DM halo when considering heavy binaries that can be seen with LISA, AEDGE and BBO. For DM masses between $10^{-13}$\,eV up to $10^{-16}$\,eV the halo mass function is bumped at scales that can be probed with these observatories (see Fig.~\ref{fig:amp_nfw} and the left panel of Fig.~\ref{fig:hmf}). In Fig.~\ref{fig:axion_bump} we show the lensing probability for BBO for the halo mass function containing both the CDM halos and the bump from miniclusters for three different DM masses $m_a$. By comparing to Fig.~\ref{fig:Pnfw}, we see that the minicluster bump in the halo mass function increases the lensing probability by $\mathcal{O}(1)$ factor.

\begin{figure}
    \centering
    \includegraphics[width=0.85\textwidth]{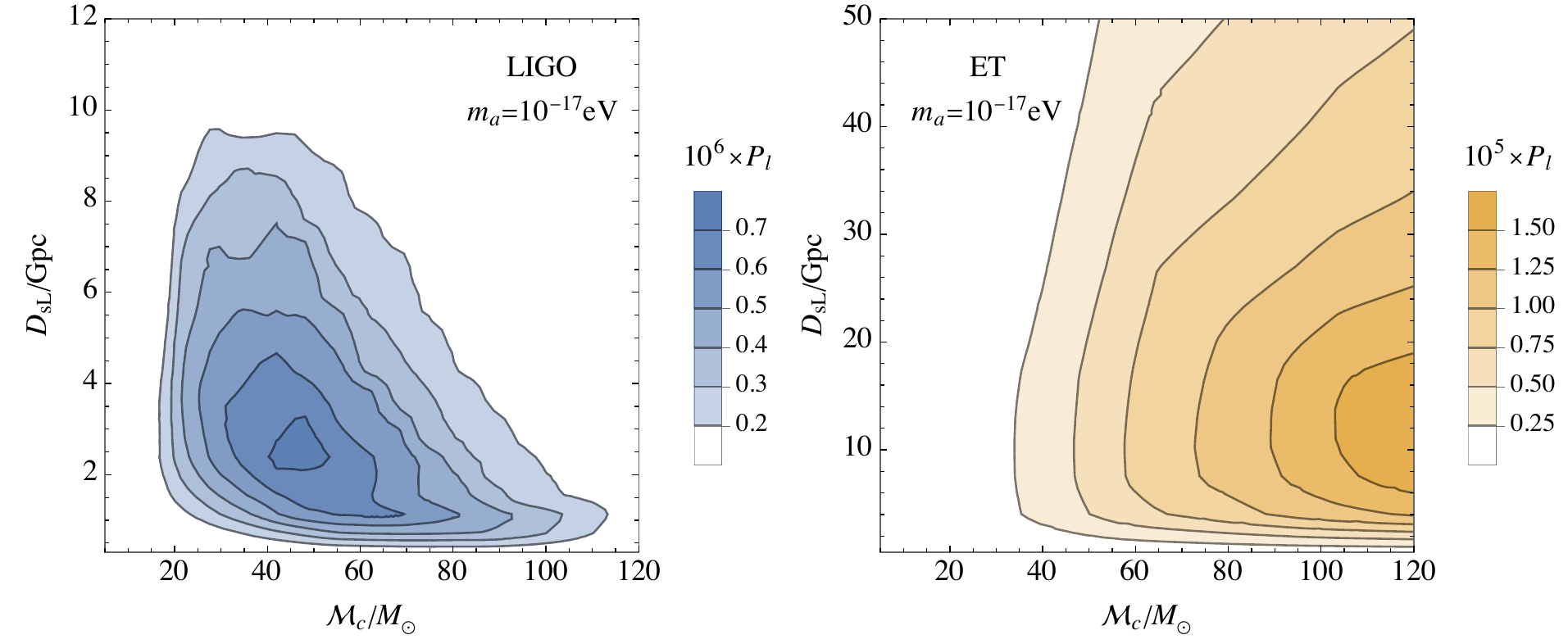}
    \caption{Probability to detect the lensing by fuzzy DM cores for $m_a=10^{-17}$\,eV as a function of the source parameters.}
    \label{fig:LIGO_ET_CDM}
\end{figure}

We focus on LIGO and ET to detect the cores inside the fuzzy DM halos since they are sensitive to the masses where the cores may be more abundant. The effect of the halo surrounding the core is negligible since for the frequencies of LIGO and ET the effect is in the geometric limit and the constant magnification of the signal would bias the luminosity distance by less than $10^{-3}$ (see Fig.~\ref{fig:amp_nfw}). The lighter $m_a$ is, the heavier the cores are per halo so there is more abundance of heavy cores. In Fig.~\ref{fig:ymaxMR} we can see that for $m_a < 10^{-18}$\,eV the cores are not compact enough to be detectable. We show in Fig.~\ref{fig:LIGO_ET_CDM} the probability to detect the cores for $m_a = 10^{-17}$\,eV, both with LIGO and ET. Roughly, this DM mass leads to the biggest microlensing probability because if DM mass is lighter, then the cores are not compact enough and if it is heavier then the lensing quickly enters the regime where the interference condition is violated.

We have a good understanding of the merger rates of the BBH that LIGO and ET may detect. To estimate the merger rate, we use the results from the most recent LIGO-Virgo catalog~\cite{LIGOScientific:2021psn} and assume a redshift dependence that follows the star formation rate~\cite{Belczynski:2016obo},
\be \label{SFR}
    {\rm SFR}(z) \propto \frac{\left(1+z\right)^{2.7}}{1+\left[(1+z)/2.9\right]^{5.6}} \,,
\ee
which 
%, as shown in the left panel of Fig.~\ref{fig:LIGO_pop}, 
is in excellent agreement with the data. 
%The chirp mass dependence of the merger rate is shown in the right panel of Fig.~\ref{fig:LIGO_pop}. 
The expected number of detectable lensed events for $m_a=10^{-17}\rm eV$ is $0.001^{+0.003}_{-0.0005}$ for LIGO and $0.03_{-0.02}^{+0.04}$ for ET during a year of observation. The uncertainties in the expected number of detectable lensed events reflect the uncertainties in the LIGO-Virgo merger rate.
% shown in the right panel of Fig.~\ref{fig:LIGO_pop}.

%\begin{figure}
%     \centering
%     \includegraphics[width=0.95\textwidth]{LIGO_pop.pdf}
%     \caption{\emph{Left panel:} Comparison of the redshift dependence from Eq.~\eqref{SFR} (solid curve) with the O3 results of LIGO-Virgo from~\cite{LIGOScientific:2021psn} (gray band). \emph{Right panel:} The merger rate of BBH as a function of the chirp mass $\mathcal{M}$ from~\cite{LIGOScientific:2021psn}.}
%     \label{fig:LIGO_pop}
%\end{figure}

\subsection{Dark matter stars}

In order to demonstrate The reason why the expected number of lensed events in the above cases is small is that there isn't sufficiently many lens halos. To demonstrate that an increase in the abundance of lens objects indeed leads to large number of detectable lensed events, we consider a scenario where fraction $f_{\rm DM}$ of all DM is in objects that can be approximated by a constant density sphere profile, that is, a fraction of DM constitutes DM stars. 

\begin{figure}
     \centering
     \includegraphics[width=0.95\textwidth]{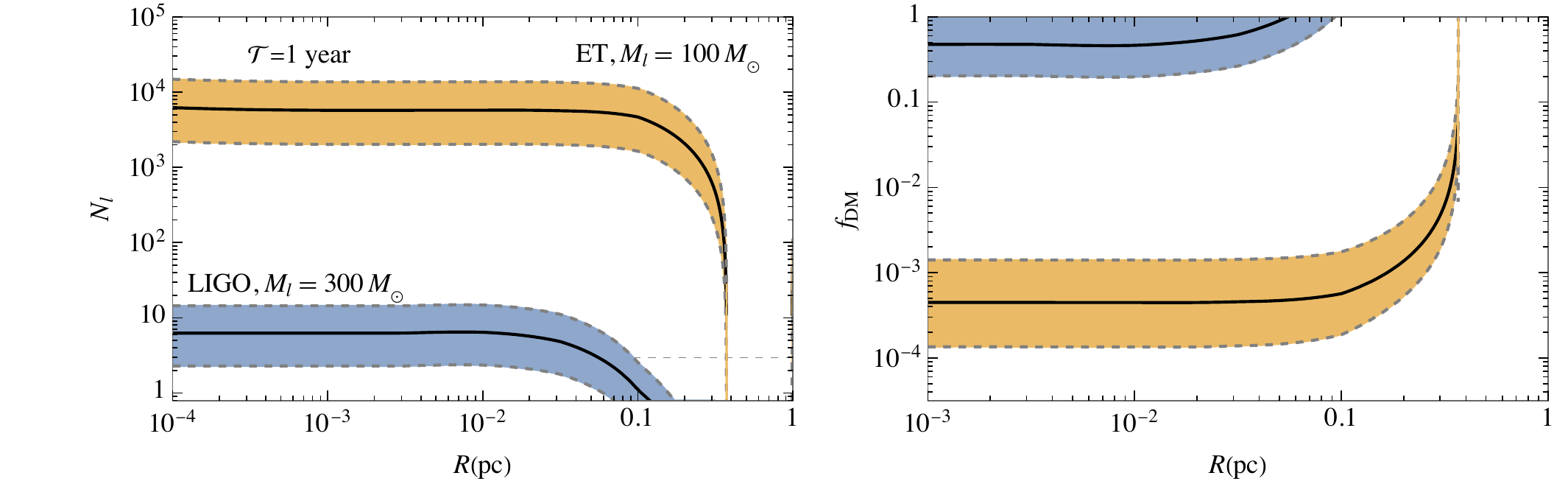}
     \caption{\emph{Left panel:} Expected number of lensed events assuming that all DM is in the form of compact objects of radius $R$ and mass $M_{\rm l}=100 M_{\odot}$ during a year of observation with ET (yellow) or with LIGO (blue). The error bands reflect the uncertainties in the BH merger rate. \emph{Right panel:} Projected constraints from the non-observation of microlensing signatures.}
     \label{fig:Rconstr}
\end{figure}

From Fig.~\ref{fig:ymaxMR}, we see that the best case to detect the microlensing effect induced by such objects is at $M_l \simeq 100 M_\odot$ for ET and at $M_l \simeq 300M_\odot$ for LIGO. Taking these values as benchmark cases, we show in the left panel of Fig.~\ref{fig:Rconstr} the expected number of detectable microlensed events assuming they constitute all DM, $f_{\rm DM}=1$, as a function of their radius $R$. Here we have used again the merger rate indicated by the LIGO-Virgo observations. We see that if these objects are sufficiently compact LIGO will in this case see $\mathcal{O}(1)$ lensed event per year and ET $\mathcal{O}(1000)$ lensed events per year.\footnote{Notice that, our analysis includes only the inspiral phase, whereas, in particular for LIGO, the merger and ringdown phases are important and including them would increase the expected number of detectable events. We leave a study going beyond the inspiral phase for future work.} The finite size of the lens becomes relevant at $R > 0.01$\,pc for LIGO and at $R > 0.1$\,pc for ET, and the expected number of lensed events quickly falls to zero. In the right panel of Fig.~\ref{fig:Rconstr} we show the projected sensitivity of these experiments on the fraction of DM in these DM stars as a function of their radius. If ET will not see any events, it will be able to put a strong constraint, $f_{\rm DM} \leq 10^{-3}$, on the abundance of such objects if their radius is $R < 0.37$\,pc. The projected constraint from LIGO in its design sensitivity may reach $f_{\rm DM} \simeq 0.5$, depending on the BH merger rate.

\section{Conclusions}

In this work, we have considered microlensing of GWs from BH coalescence events by small dark halos along the path of the wave as a probe of small scale DM structures. We have shown that for the consistency of the microlensing formalism, the classical paths that the GW takes around the DM structure should get to the detector at approximately the same frequency. This imposes constraints on the values of the impact parameter $y$ and the lens mass $M_{lz}$, as shown in Fig.~\ref{fig:consistency} when we impose the microlensing interference condition up to the ISCO orbit. For the microlensing effect of the lens to be relevant, we have also shown in Fig.~\ref{fig:consistency} that the mass lens cannot be much lighter than the chirp mass of the binary $\mathcal{M}$, as otherwise, the classical paths do not produce any interference effect at the detector. The combination of both effects imposes a range of lens masses for a given $y$ and $\mathcal{M}$ where the microlensing formalism describes an interference effect that could be detectable. The interference can happen in the geometric limit or in the wave optics regime. 
%The former correctly describes the interaction when the time of arrival oscillates quickly enough in the lens plane except for around the classical paths (see Fig.~\ref{fig:oscillations}). This roughly happens when the wavelength of the GW is smaller than the Schwarzschild radius of the lens. 
The former correctly describes the effect roughly when the wavelength of the GW is smaller than the Schwarzschild radius of the lens.

First, we have considered microlensing by CDM halos. This search is relevant for halos lighter than $10^8 M_{\odot}$, where the number of stars is small enough to challenge conventional techniques. We have approximated the CDM halos by NFW profiles, which we have checked to be close enough to the Einasto one that is slightly preferred from numerical simulations. Such light spherically symmetric DM halos are not compact enough to make more than one classical path around them and in this case only the wave effects around the shortest path may produce a detectable effect. We have checked that only AEDGE, LISA and BBO are going to be sensitive to the wave effect for halos of virial mass in the range from $10^4M_\odot$ to $10^8M_\odot$, see Fig.~\ref{fig:amp_nfw}. For these detectors, we have computed, assuming the CDM halo mass function, what is the probability for an event to be lensed. We have shown in Fig.~\ref{fig:Pnfw} that the maximal probabilities for LISA, AEDGE and BBO are, respectively $5\times 10^{-5}$, $10^{-5}$, and $1.3\times10^{-3}$. There are big uncertainties in the merger rate of IMBHs that these observatories may detect. We show that for current estimates, even for BBO the number of lensed events a year is small, $\mathcal{O}(0.01/{\rm year})$.

Second, we have considered fuzzy DM halos, which are expected to host solitonic cores that we have approximated by uniform density sphere profiles. Although it is not possible to find analytical expressions for the microlensing effect of the cores, we have found three main lensing types depending on how many classical paths exist outside the sphere. The different regimes are shown in Fig.~\ref{fig:AmpDMS}. The finite size effects are relevant for the interaction with the cores if their radius is larger or comparable to the Einstein radius. In particular, we have shown in Fig.~\ref{fig:ymaxMR} that ET can probe the microlensing effect for lenses with sizes up to an order magnitude bigger than their Einstein radius. The relevant fuzzy DM masses for detecting the microlensing effect by LIGO or ET range roughly from $10^{-18}$\,eV to $10^{-16}$\,eV. We have shown that the maximal probability to detect the microlensing is $10^{-6}$ for LIGO and $3\times 10^{-5}$ for ET, realized for DM mass $m_a\simeq 10^{-17}\rm eV$. Using the merger rate of $\mathcal{O}(10M_\odot)$ BHs indicated by the LIGO-Virgo observations and assuming that the redshift dependence follows the star formation rate, we have estimated the number of events lensed by the cores. For the best case DM mass, the expected number of lensed events is $\mathcal{O}(0.01/{\rm year})$. If fuzzy DM is in the form of axion-like particles and the PQ symmetry was broken before matter-radiation equality, we expect to see deviations from the CDM halo mass function since there are contributions also of axion miniclusters. This, however, induces only an $\mathcal{O}(1)$ increase in the expected number of lensed events.

Third, we have considered microlensing by DM stars. We have shown that, assuming their mass is in the range detectable with LIGO or ET (see Fig.~\ref{fig:ymaxMR}), the expected number of detectable microlensed events can be large, $\mathcal{O}(10^3)$ for ET and $\mathcal{O}(1)$ for LIGO. The finite radius of these stars becomes relevant for ET at $R>0.1$\,pc and for LIGO at $R>0.01$\,pc, and the microlensing effects become undetectable if the radius is significantly larger than that. As shown in Fig.~\ref{fig:Rconstr}, the large expected number of lensed events turns into a strong constraint on the abundance of these objects, in case such microlensing events will not be observed.

This work shows that probing dark structures through microlensing of GWs is difficult. With developments in future sensitivity, it may be possible to discover compact dark structures, and DM scenarios that predict compact objects in the stellar mass range can lead to predictions using this methodology. In the case of relatively compact DM objects, such as fuzzy DM cores, the lensing probability is limited by imposing the microlensing interference condition. This implies that accounting for the strong lensing regime may help in detecting such objects. We leave the study of the detectability of the lens effects ranging from microlensing to strong lensing due to compact DM structures for future work. Moreover, whereas the probability of seeing any individual events lensed by NFW halos is very low, it may be possible to see the lensing effects statistically, in the way of weak lensing, by studying the whole population of gravitational wave events. We leave also such studies for future work.

\section*{Acknowledgements}
The authors are extremely grateful for conversations with Doddy Marsh.  This work was supported by European Regional Development Fund through the CoE program grant TK133 and by the Estonian Research Council grant PRG803. The work of VV has been partially supported by the European Union's Horizon Europe research and innovation program under the Marie Sk\l{}odowska-Curie grant agreement No. 101065736.  MF gratefully received funding via the STFC particle theory grant STFC-ST/T000759/1.

\bibliographystyle{JHEP}
\bibliography{refs}

\end{document}